\documentclass[twocolumn,letterpaper,amsmath,amssymb,floatfix,aps,superscriptaddress]{revtex4-1}

\usepackage{hyperref}
\usepackage{verbatim}
\usepackage{graphicx}
\usepackage{bm} 
\usepackage{ifpdf} 
\usepackage{dcolumn}  
\usepackage{epsfig}
\usepackage{color}
\usepackage[usenames,dvipsnames]{xcolor}

\newcommand{\Av}[1]{{\bf #1}} 
\newcommand{\Avg}[1]{{\boldsymbol #1}}

\def\rmd{{\mathrm{d}}}

\def\rme{{\mathrm{e}}}
\def\lB{l_{\mathrm{B}}}

\begin{document}

\title{Ion-mediated interactions between net-neutral slabs: Weak and strong disorder effects}


\author{Malihe Ghodrat}
\affiliation{School of Physics, Institute for Research in Fundamental Sciences (IPM), Tehran 19395-5531, Iran}

\author{Ali Naji}
\thanks{Email: \texttt{a.naji@ipm.ir}}
\affiliation{School of Physics, Institute for Research in Fundamental Sciences (IPM), Tehran 19395-5531, Iran}

\author{Haniyeh Komaie-Moghaddam}
\affiliation{School of Physics, Institute for Research in Fundamental Sciences (IPM), Tehran 19395-5531, Iran}

\author{Rudolf Podgornik}
\affiliation{Department of Theoretical Physics, J. Stefan Institute, SI-1000 Ljubljana, Slovenia}
\affiliation{Department of Physics, Faculty of Mathematics and Physics, University of Ljubljana, SI-1000 Ljubljana, Slovenia}


\begin{abstract}
We investigate the effective interaction between two randomly charged but otherwise net-neutral, planar dielectric slabs immersed in an asymmetric Coulomb fluid containing a mixture of mobile monovalent and multivalent ions. The presence of charge disorder on the apposed bounding surfaces of the slabs leads to substantial qualitative changes in the way they interact, as compared with the standard picture provided by the van der Waals and image-induced, ion-depletion interactions. While, the latter predict purely attractive interactions between strictly neutral slabs, we show that the combined effects from surface charge disorder, image depletion, Debye (or salt) screening and also, in particular, their coupling with multivalent ions, give rise to a more diverse behavior for the effective interaction between net-neutral slabs. Disorder effects show large variation depending on the properly quantified strength of disorder, leading either to non-monotonic effective interaction with both repulsive and attractive branches when the surface charges are weakly disordered (small disorder variance) or to a dominating attractive interaction that is larger both in its range and magnitude than what is predicted from the van der Waals and image-induced, ion-depletion interactions, when the surfaces are strongly disordered (large disorder variance).
\end{abstract}


\maketitle
\section{Introduction}
 
Interactions between neutral dielectric bodies are traditionally viewed as being due predominantly to electromagnetic field fluctuations, or equivalently, dipole fluctuations that give rise to van der Waals (vdW) fluctuation-induced interactions between them \cite{Parsegian2005,Ninham76}. These interactions are attractive between identical bodies in vacuum or in a polarizable medium, such as water or an aqueous electrolyte (or Coulomb fluid). They contribute one of  the two key ingredients of the classical Derjaguin-Landau-Verwey-Overbeek (DLVO)  theory of colloidal stability, the other one being the mean-field electrostatic interaction, which is repulsive for like-charged colloidal surfaces \cite{VO,Israelachvili,andelman-rev,Safranbook,holm}.

Recent works have, however, highlighted the role of image-induced, ion-depletion effects in this scenario (see, e.g., Refs. \cite{Hansen,Dean04,Zemb,Kjellander87,Kjellander08,Netz1,Netz2,Netz3,Monica,Monica2,Monica3,Wang0,Wang1,Wang2,Levin1,Levin2,Levin2b,Levin3,Levin4,Lue0,Lue1,Lue1b,Lue2,Markovich1,Markovich2,Sahin0,Sahin5,Sahin2,Sahin3,Sahin4,SCdressed3} and references therein), leading to depletion of mobile solution ions from the vicinity of  dielectric interfaces and, therefore, to an additional attractive force between apposed dielectric boundaries. This is because most dielectric surfaces in the context of bio- and soft materials have a lower (static) dielectric constant than that of water and, therefore, solution ions experience repulsion from their same-sign image charges in the proximity of  dielectric boundaries \cite{Honig}. The recent advances in the study of image-induced, ion-depletion effects follow on the trail of the Onsager-Samaras framework formulated originally in the context of the surface tension problem of electrolytes \cite{Wagner,OS}. Such non-mean-field effects, which belong to the general class of depletion interactions \cite{lekker}, arise due to the discrete nature of mobile ions neglected in the collective mean-field description based on the standard Poisson-Boltzmann theory \cite{VO,Israelachvili,andelman-rev,Safranbook,holm}. The studies of image-induced, ion-depletion effects have been focused exclusively on the case of strictly neutral (charge-free) dielectric surfaces. In this case, the ion-depletion interactions can be amplified in the presence of mobile multivalent ions in the solution \cite{SCdressed3,Kjellander08} due to stronger ion-image repulsions for these ions, even when the multivalent ions are present at small bulk concentrations around just a few mM \cite{SCdressed3}. 

In this paper, we revisit the problem of interaction between neutral dielectric surfaces in a Coulomb fluid by adding to it a novel feature: We relax the constraint of {\em strict} electroneutrality of surface boundaries, considered so far in the literature, by assuming that the surfaces are neutral only {\em on the average}, while microscopically they carry a {\em quenched} (fixed) random distribution of positive and negative charges. We show that this seemingly simple generalization leads to significant qualitative changes in the distribution of ions and, consequently, also in the effective interactions between dielectric surfaces, especially when the intervening Coulomb fluid contains mobile  multivalent ions.

Disordered charged systems are abundant  in soft matter with examples ranging from polycrystalline surfaces with patchy surface potentials \cite{kim2,kim3,barrett}, dielectric contact surfaces \cite{science11}, vapor-deposited amorphous films on solid substrates \cite{liu}, surfactant-coated surfaces \cite{Meyer,Meyer2,klein,klein1,klein2}, DNA microarrays \cite{science95,Zhulina04}, intrinsically disordered proteins \cite{Uversky,Artem}, patchy colloids \cite{likos} and random polyelectrolytes and polyampholytes \cite{kantor-disorder0,andelman-disorder,Bing}. Surface charge disorder in these examples can exhibit highly random distributions as well as patchy and heterogeneous patterns, originating from different sources including specific electronic and/or structural properties of materials involved, surface grafting or adsorption of charged molecules and/or contaminants, synthetic and fabrication processes, etc. The surface charge disorder can be highly sample specific and, at the same time, can depend  strongly on the method of preparation. It may be set and quenched for each sample (which is the case of interest in this paper), annealed (in which case the surface charges are mobile and in thermal equilibrium with the  rest of the system), or partially quenched or partially annealed \cite{partial,Hribar}. Motivated by these examples, study of charge disorder and, in particular, effective interactions between random charge distributions has witnessed growing attention from theoreticians over the last several years \cite{speake,Bing,klein2,kantor-disorder0,andelman-disorder,Miklavic,Ben-Yaakov-dis,netz-disorder,netz-disorder2,Lukatsky1,Lukatsky2,Rabin,safran1,safran2,safran3,Olvera0,Hribar,ali-rudi,rudiali,partial,disorder-PRL,jcp2010,pre2011,epje2012,jcp2012,book,jcp2014,sm2015}, as well as from the simulation side where initial steps have been taken to include the effects of charge heterogeneity and disorder \cite{Amin,jho06,netz-disorder,netz-disorder2,safran3,Olvera0,jho12}. Disorder effects have been studied extensively in the context of electromagnetic fluctuation-induced interactions between two apposed, randomly charged surfaces in vacuum \cite{disorder-PRL,jcp2010,pre2011,epje2012,jcp2012,book,speake,kim3}, where they have been associated with anomalously long-ranged surface interactions observed in recent experiments \cite{kim2,kim3}. They have also been studied in situations where a weakly coupled Coulomb fluid, containing, e.g., monovalent cation and anions, intervenes between the bounding surfaces \cite{klein2,Hribar,Miklavic,Ben-Yaakov-dis,netz-disorder,netz-disorder2,Rabin,safran1,safran2,safran3,ali-rudi,rudiali,partial,pre2011,Olvera0}. This however leaves out the case of asymmetric Coulomb fluids containing both monovalent ions and multivalent (counter-) ions. These kinds of systems are in fact quite common in experiments within  the biological context as in the case of viruses, DNA condensates or other charged biopolymer aggregates  \cite{rau-1,Angelini03,Needleman,Bloom2,Yoshikawa2,Pelta,Plum,Raspaud,Savithri1987,deFrutos2005,Siber} and are expected to behave very differently since multivalent ions are known to couple strongly with fixed surface charges.

Strong-coupling behavior of multivalent counterions at uniformly charged surfaces or surfaces with regular charge patterns has been studied extensively over the last decade and its connection to exotic phenomena such as like-charged attraction has been throughly discussed (for recent reviews and a more exhaustive list of references, see  Refs. \cite{holm,book,Levin02,Shklovs02,hoda_review,Naji_PhysicaA,French-RMP,perspective}). In disordered charged systems, such strong-coupling phenomena have been considered only in a few cases so far \cite{ali-rudi,partial,jcp2014,sm2015} by assuming that bounding surfaces carry a {\em finite} mean surface charge density, to which multivalent counterions can couple strongly, in the same sense as considered in the context of uniformly charged surfaces as noted above. Yet, the presence of charge randomness on bounding surfaces was shown to give rise to novel phenomena such as strong surface attraction of multivalent counterions, characterized by a density profile that diverges at the surface, in clear violation of the contact-value theorem established for uniformly charged surfaces.  This kind of behavior originates from a singular, attractive, single-ion potential, which is created by the randomness in the distribution of surface charges and, as such, depends on the surface charge variance. Consequently, one can also show that the thermal entropy of counterions is diminished upon introducing a finite degree of charge randomness on the boundaries. In other words, the system becomes more `ordered' as a direct outcome of the interplay between the thermal entropy of ions and the configurational entropy of charge randomness, entering through an ensemble average over various realizations of the disordered charge distribution. This peculiar disorder-induced effect, which stands at odds with what one may expect intuitively, has been referred to as {\em antifragility} \cite{taleb}.

Our analysis in this paper is focused on yet another facet of the disorder-induced effects by assuming, in a system of two plane-parallel dielectric slabs, that the randomly charged surfaces of the slabs bear {\em no net charge}. This eliminates the direct strong coupling between multivalent ions and the mean surface charge and, thereby, also the ensuing strong-coupling interactions, considered in our previous papers  \cite{jcp2014,sm2015}, that would otherwise completely mask the vdW and image-induced, ion-depletion interactions between the slabs in an asymmetric Coulomb fluid. This allows us to address the question of how the presence of surface charge disorder affects the standard picture for the interaction of neutral bodies based on the vdW and ion-depletion effects.

The surface charge disorder has several different implications: First, it contributes a {\em repulsive} interaction between the slabs, which comes from self-interactions of disorder charges on the bounding surfaces with their image charges; this contribution counteracts the {\em attractive} vdW interaction as discussed thoroughly elsewhere \cite{rudiali,disorder-PRL,jcp2010,pre2011,epje2012,jcp2012,book}. Then, as noted above, the individual ions in the Coulomb fluid experience an attractive disorder-induced potential, which strongly attracts them toward the bounding surfaces. This effect counteracts the ion-image repulsions that tend to deplete ions from the slit region between the slabs and form the basis of the image-induced, ion-depletion mechanism for attraction between strictly neutral slabs. Thus, randomly charged bounding surfaces tend to accumulate more mono- and multivalent ions in the slit. The system, however, responds differently to the increased number of ions: While the osmotic pressure due to monovalent ions increases and even becomes {\em repulsive}, consistent with the standard mean-field picture, the osmotic pressure due to multivalent ions, becomes ever more {\em attractive}! This behavior is rooted in the combined effect of the surface charge disorder and the presence of mobile multivalent ions, with the latter creating strong inter-surface attractions upon further accumulation in the slit. As a result, the effective total interaction pressure between randomly charged, net-neutral dielectric surfaces can differ qualitatively from what one expects based on the standard vdW \cite{Parsegian2005,Ninham76} and image-induced, ion-depletion interactions in the case of strictly neutral surfaces  (see, e.g., Refs. \cite{Hansen,Dean04,Zemb,Kjellander87,Kjellander08,Netz1,Netz2,Netz3,Monica,Monica2,Monica3,Wang0,Wang1,Wang2,Levin1,Levin2,Levin2b,Levin3,Levin4,Lue0,Lue1,Lue1b,Lue2,Markovich1,Markovich2,Sahin0,Sahin5,Sahin2,Sahin3,Sahin4,SCdressed3} and references therein).

The net effect due to the interplay between disorder, mono- and multivalent-ion contributions yields a qualitatively different behavior for the effective surface-surface interactions, depending on the strength of disorder as quantified by the {\em disorder coupling (or strength) parameter} \cite{ali-rudi,partial,jcp2014,sm2015}. This feature bears some conceptual resemblance to the strong-weak coupling dichotomy that exists for net-charged surfaces, dependent on the {\em electrostatic coupling parameter} in that case \cite{Netz01,AndrePRL,AndreEPJE}. In weakly disordered systems, corresponding to a small disorder coupling parameter, one then discerns a distinct non-monotonic behavior for the interaction pressure as a function of the separation between the slabs, with a pronounced repulsive hump at intermediate separations; conversely, in strongly disordered systems, corresponding to a large disorder coupling parameter, the effective interaction pressure becomes strongly attractive, with a range and magnitude larger than that of the vdW or the image-induced, ion-depletion interaction pressure as found between strictly neutral surfaces. 

The organization of the paper is as follows: In Section \ref{sec:model}, we introduce our model and, in Section \ref{sec:general}, we briefly discuss the theoretical background and present the general results for the two-slab system. The results of our analysis for the density profile of multivalent ions and the effective interactions between the slabs are presented in Section \ref{sec:results}. We conclude our discussion  in Section \ref{sec:conclusion}.

\begin{figure}[t]
 \centering
 \includegraphics[width=8.5cm]{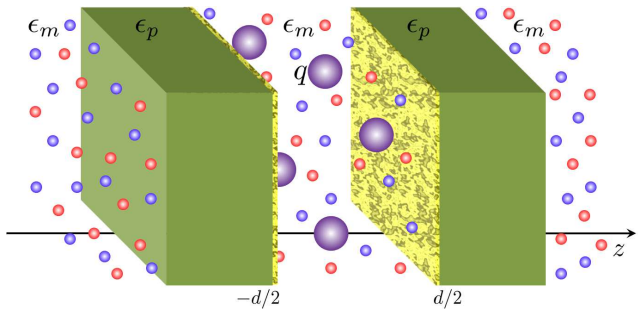}
 \caption{(Color online) Two infinite, plane-parallel dielectric slabs of dielectric constant $\epsilon_p$ and randomly charged inner surfaces are immersed in a bathing ionic solution of dielectric constant $\epsilon_m$. The solution contains a mixture of monovalent and multivalent salts. Multivalent ions are confined in the slit region and are shown by large spheres; monovalent salt anions and cations are shown by small red and blue spheres. The slabs are assumed to be neutral on the average and their thickness is taken to be infinite.}
 \label{f:schematic}
\end{figure}

\section{The model}
\label{sec:model}

Our model consists of two plane-parallel dielectric slabs of infinite surface area $S$ and dielectric constant $\epsilon_p$ with inner surfaces placed perpendicular to the $z$ axis at a separation distance of $d$ (see Fig.~\ref{f:schematic}). The slabs are immersed in an asymmetric Coulomb fluid of dielectric constant $\epsilon_m$ at room temperature $T$. The Coulomb fluid is a mixture of a monovalent $1:1$ salt of bulk (reservoir) concentration $n_0$ and a multivalent $q:1$ salt of bulk concentration $c_0$ with multivalent ions having charge valency of $q>0$. The multivalent ions are confined within the slit region $-d/2 \leq z \leq d/2$, while the monovalent ions are dispersed throughout the space except in the region occupied by the dielectric slabs that are assumed to be impermeable to mobile ions \cite{Note}. We are interested only in the cases where the slab thickness is much larger than the Debye (or salt) screening length and, thus, in the calculations to be presented later, we shall assume that the slab thickness is infinite.

The inner surfaces of the slabs are assumed to bear a {\em quenched}, random charge distribution $\rho(\Av r)$, while they remain electrically {\em net-neutral}. We assume that the random surface charges are distributed according to a Gaussian probability distribution function, which is determined fully by its two moments, $ \langle\!\langle\rho(\Av r)\rangle\!\rangle=0$, and
\begin{equation}
  \langle\!\langle\rho(\Av r)\rho(\Av r')\rangle\!\rangle= g(\Av r)\delta(\Av r-\Av r').
\end{equation}
with
\begin{equation}
g({\Av r})= g e_0^2 \, \big[\delta(z+d/2)+\delta(z-d/2)\big],
\label{eq:g}								
\end{equation}
where $g\geq 0$ is the surface disorder variance. By assumption, thus, the disordered charge distributions on the two slabs are statistically uncorrelated. The effects of surface charge correlations (or ``patchiness") and the internal structure of multivalent ions, which are assumed to be point-like here, will be considered elsewhere \cite{preprint4,preprint5}. 

\section{Interaction free energy: General results}
\label{sec:general}

\subsection{Formal background}

In the present model, we have two types of mobile ions, namely, mono- and multivalent ions, that are expected to behave very differently in the presence of dielectric boundaries. On general grounds, and if the surfaces are assumed to be strictly neutral (charge free), mobile ions interact only with their image charges. This interaction scales with the second power of the ionic valency, being thus much larger in the case of multivalent ions with $q\gg 1$ (e.g., trivalent and tetravalent ions). In this latter case, 
the ion-image interactions are dominated by the self-image interaction. 
Note that in most realistic examples, ions are dissolved in water, which is a highly polarizable medium with  $\epsilon_m=80$ at room temperature ($T=293$~K) and, thus, we often have $\epsilon_m>\epsilon_p$, giving image charges of the same sign and, hence, repulsive self-image interactions for each ion in the slit. Such repulsive interactions lead to statistical correlations (manifested as ion depletion) between the mobile ions and the bounding surfaces that are expected to be weak for monovalent ions, but quite strong for multivalent ions. This leads to a complicated problem in which different ionic species show distinctly different couplings to local electrostatic fields generated by the boundaries. While the monovalent ions can be therefore described appropriately by mean-field-type theories, such as the Poisson-Boltzmann or the Debye-H\"uckel theory \cite{VO,Israelachvili,Safranbook,holm,andelman-rev}, the multivalent ions require an altogether different description that should account for such large correlation effects.

The situation is, therefore, analogous to the one found in the case of asymmetric Coulomb fluids confined between {\em charged} surfaces, where the dominant factor determining the behavior of multivalent counterions is again their electrostatic coupling to the local fields that are generated by the boundaries. The difference is however that in the the case of a charged surface the coupling is determined by the so-called electrostatic coupling parameter \cite{Netz01,AndrePRL,AndreEPJE}, depending on the charge density of the surface boundaries, enabling the so-called {\em strong-coupling approximation} for multivalent counterions at charged surfaces (see, e.g.,  Refs. \cite{holm,hoda_review, perspective, Naji_PhysicaA, Levin02, Shklovs02,French-RMP,book,Netz01,AndrePRL,AndreEPJE,asim,Naji_epje04,Naji_epl04,jho-prl,Weeks,Weeks2,Burak04,trizac} and references therein), or more generally, the {\em dressed multivalent-ion theory} \cite{SCdressed1,SCdressed2,SCdressed3,perspective}, which provides a very good approximation for the study of asymmetric Coulomb fluids over a wide range of parameters as verified by explicit-ion and implicit-ion simulations. The dressed multivalent-ion theory  reproduces  the mean-field Debye-H\"uckel theory and the standard  strong-coupling theory for counterion-only systems \cite{hoda_review, perspective, Naji_PhysicaA,Levin02, Shklovs02,Netz01,AndrePRL,AndreEPJE,asim,Naji_epje04,Naji_epl04,jho-prl,Weeks,Weeks2,Burak04}  as two limiting theories in the regime of large and small Debye screening lengths and, therefore, bridges the gap between these two limits \cite{SCdressed1,SCdressed2,SCdressed3,perspective,leili1,leili2}. The key step in this latter approach is to integrate out the degrees of freedom associated with monovalent ions by means of a linearization approximation, justified only for highly asymmetric Coulomb fluids $q\gg 1$, and yielding an effective  Debye-H\"uckel (DH) interaction between the remaining multivalent ions and the surface charges (if any) \cite{SCdressed1}. Then, one expands the partition function of the system in terms of the fugacity (or bulk concentration, $\lambda_c=c_0$) of multivalent ions or in terms of the inverse electrostatic coupling parameter for counterion-only systems \cite{Netz01}. The leading order of the virial expansion can be cast into a simple analytic theory because of its single-particle structure, which can successfully describe various features of strongly charged systems containing multivalent ions (see, e.g. Refs. \cite{hoda_review, perspective, Naji_PhysicaA,Netz01,AndrePRL,AndreEPJE,asim,Naji_epje04,Naji_epl04,jho-prl,Weeks,Weeks2,Burak04,SCdressed1,SCdressed2,SCdressed3,trizac,Shklovs02,holm,book,leili1,leili2,Levin02} and references therein). The regime of applicability of this theory has been discussed extensively in recent literature \cite{SCdressed1,SCdressed2,SCdressed3,hoda_review, perspective,Naji_PhysicaA,AndreEPJE,trizac,jcp2014,sm2015,Weeks,Weeks2,Burak04,holm,asim,Naji_epje04,Naji_epl04,jho-prl}, which we therefore do not reiterate here.

In the case of strictly neutral (charge-free) surfaces, this strategy was shown to be effective as well \cite{SCdressed3}, since, as is often the case in experimental systems containing asymmetric ionic mixtures \cite{Bloom2, Yoshikawa2,Pelta, Plum,Raspaud,Savithri1987,deFrutos2005,Siber,rau-1,Angelini03,Needleman},  multivalent counterions of high valency (for instance, CoHex$^{3+}$ or polyamines such as Spd$^{3+}$, Sp$^{4+}$) are present only in small bulk concentrations, of about a few mM, justifying fully the virial expansion scheme that underlies the dressed multivalent-ion theory. This is the approach of choice that we adopt also in the present context, where the surface boundaries carry random charges and are neutral only on the average. The only additional step here is that, due to the quenched disorder in the surface charge distribution, the free energy of the system follows by averaging over the whole ensemble of charge distributions, $\rho(\Av r)$ \cite{jcp2014,sm2015}.

\subsection{Planar slabs}

Without delving further into details available in previous publications \cite{SCdressed1,SCdressed2,SCdressed3,perspective,jcp2014,sm2015}, we proceed by giving the general expression for the free energy of the two-slab system considered in this work, which has the form \cite{sm2015}
\begin{equation}
\label{eq:exact_Gpot}
 {\mathcal  F}=\frac{1}{2}\mathrm{Tr}\big[g(\Av r)G(\Av r,\Av r')\big]- \lambda_c k_{\mathrm{B}}T\int {\mathrm{d}}\Av r\, \Omega(\Av r)\, \rme^{-\beta u(\Av r)}.
\end{equation}
Here, $\beta=1/(k_{\mathrm{B}}T)$ and the Green's function of the system is defined via
\begin{equation}
-\epsilon_0 \nabla\cdot \epsilon({\mathbf r}) \nabla G({\mathbf r}, {\mathbf r}')  + \epsilon_0  \epsilon({\mathbf r})\kappa^2({\mathbf r})G({\mathbf r}, {\mathbf r}')= \delta({\mathbf r} -{\mathbf r}'),
\label{eq:G_DH}
\end{equation}
where $\kappa({\mathbf r})$ is the Debye (or salt) screening parameter, which is non-zero only outside the region occupied by the dielectric slabs, i.e., $\kappa^2 = 4\pi l_{\mathrm{B}} n_b$ with $n_b = 2n_0+qc_0$ being the total bulk concentration of monovalent ions.

The first term in Eq. (\ref{eq:exact_Gpot}) represents the free energy of the system in the absence of multivalent ions. It stems from {\em self-interactions} of disorder charges on the bounding surfaces of the slabs with their image charges  (note that the random charge distributions on the two slab are uncorrelated and, on the average, do not interact with each other). This term depends only on the disorder variance, $g$, and is non-vanishing only in inhomogeneous systems with a finite dielectric discontinuity at the bounding surfaces and/or a spatially inhomogeneous distribution of salt ions. This contribution has been analyzed in the context of the fluctuation-induced forces between disordered surfaces in vacuum or in a weakly coupled Coulomb fluid \cite{rudiali,disorder-PRL,jcp2010,pre2011,epje2012,jcp2012,book}.

The second term in Eq. (\ref{eq:exact_Gpot}) represents the contribution from multivalent ions on the leading (virial) order, in which $u({\mathbf r})$ is
the {\em effective} single-particle interaction energy \cite{jcp2014,sm2015}
\begin{equation}
 u({\mathbf r}) = \frac{q^2e_0^2}{2} G_{\mathrm{im}}({\mathbf r}, {\mathbf r})
   - \beta  \frac{q^2e_0^2}{2} \int {\mathrm{d}}{\mathbf r}' g(\Av r') [G(\Av r,\Av r')]^2.
   \label{eq:u2}
\end{equation}
The first term in the above expression represents the self-image interaction of individual multivalent ions in the slit and the second term represents the contribution of the surface charge disorder to the effective single-particle interaction energy. In the first term, $G_{\mathrm{im}}({\mathbf r}, {\mathbf r})= G({\mathbf r}, {\mathbf r}) - G_0({\mathbf r}, {\mathbf r})$ gives the dielectric and/or salt polarization, or the  image-charges effects (corresponding to the generalized Born energy), in which the free-space Green's function, $G_0({\mathbf r},{\mathbf r})$ (corresponding to the formation energy of individual ions in a homogeneous background), defined via $-\epsilon_0 \epsilon_m(\nabla^2 - \kappa^2) G_0({\mathbf r}, {\mathbf r}')  = \delta({\mathbf r} -{\mathbf r}')$, is subtracted from the total Green's function. The second term in Eq. (\ref{eq:u2}) is found to be proportional to the disorder variance and shows an explicit temperature dependence and a quadratic dependence on the Green's function and the multivalent-ion charge valency, $q$ (these latter features can be understood by noting that the disorder term indeed stems from the sample-to-sample fluctuations, or variance, of the sample-dependent single-particle interaction energy as discussed in Ref. \cite{jcp2014}).

In the strong-coupling limit or within the multivalent dressed-ion theory \cite{SCdressed1,Netz01,AndreEPJE}, the number density of multivalent counterions can be obtained  in terms of the effective single-particle interaction energy as \cite{jcp2014,sm2015}
\begin{equation}
	\label{eq:sc_density_av}
         c({\mathbf r})  =    \lambda_c \Omega({\mathbf r}) \, \rme^{-\beta u({\mathbf r})}.
\end{equation}

In the specific example of two planar slabs, we can take advantage of the translational invariance of the Green's function with respect to transverse (in-plane) direction coordinates $\Avg\rho = (x, y)$ and $\Avg\rho' = (x', y')$, since $G(\Av r,\Av r') = G(\Avg\rho, \Avg\rho';z, z')$ is only a function of  $|\Avg\rho-\Avg\rho'|$, $z$ and $z'$, and write the free energy in terms of its  Fourier-Bessel transform $\hat G( Q;z,z')$ defined through
\begin{equation}
G(\Av r,\Av r') = \int_0^\infty \frac{Q {\mathrm{d}}Q}{2\pi }\,\hat G(Q;z,z')\,J_0(Q \vert \Avg\rho -\Avg\rho'\vert).
\end{equation}
For two semi-infinite slabs, one has
\begin{eqnarray}
\label{eq:G_infinite_b}
 &&\hat G(Q;z,z')= \frac{1}{2\epsilon_0\epsilon_m \gamma }\big[\rme^{-\gamma |z-z'|}+
\frac{2\Delta_s \,\rme^{-2\gamma d}}{1-\Delta_s^2\,\rme^{-2\gamma d}}\\
\nonumber &&\qquad\qquad\times\big(\rme^{\gamma d}\cosh{\gamma(z+z')}+\Delta_s \cosh{\gamma(z-z')}\big)\big].
\end{eqnarray}
where $\gamma^2=\kappa^2+Q^2$, and
\begin{equation}
\Delta_s=\frac{\epsilon_m \gamma-\epsilon_p Q }{\epsilon_m \gamma +\epsilon_p Q}.
\end{equation}
In the absence of salt screening ($\kappa=0$), $\Delta_s$ reduces to the bare dielectric discontinuity parameter
\begin{equation}
\Delta=\frac{\epsilon_m -\epsilon_p }{\epsilon_m  +\epsilon_p },
\end{equation}
which gives a measure of the magnitude of ``dielectric-image" charges; while in a {\em dielectrically homogeneous} system, we have
\begin{equation}
\Delta_s\big|_{\epsilon_m=\epsilon_p}=\frac{ \gamma- Q }{ \gamma + Q},
\label{eq:Delta_s_homogen}
\end{equation}
which gives a measure of the salt-induced image effects, or loosely speaking, ``salt-image" charges.  Both image charge effects lead to depletion of ions from the vicinity of dielectric surfaces in a medium of relatively large polarizability, but they exhibit some fundamental differences, which have been discussed in the context of charged surfaces in our previous works \cite{jcp2014,sm2015}.

\begin{figure*}[t!]\begin{center}
	\begin{minipage}[t]{0.325\textwidth}\begin{center}
		\includegraphics[width=\textwidth]{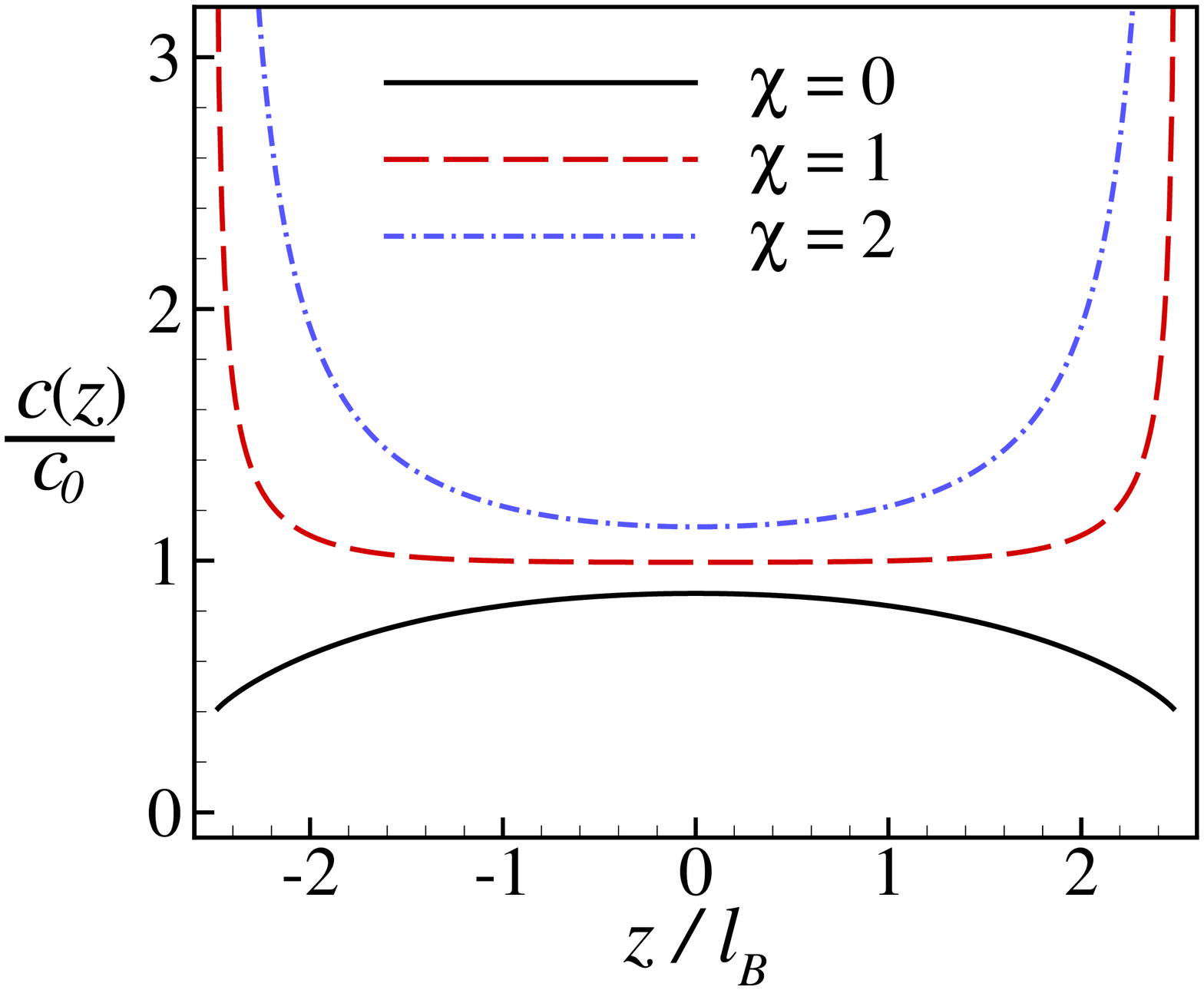} (a)
	\end{center}\end{minipage} \hskip0.1cm	
	\begin{minipage}[t]{0.325\textwidth}\begin{center}
		\includegraphics[width=\textwidth]{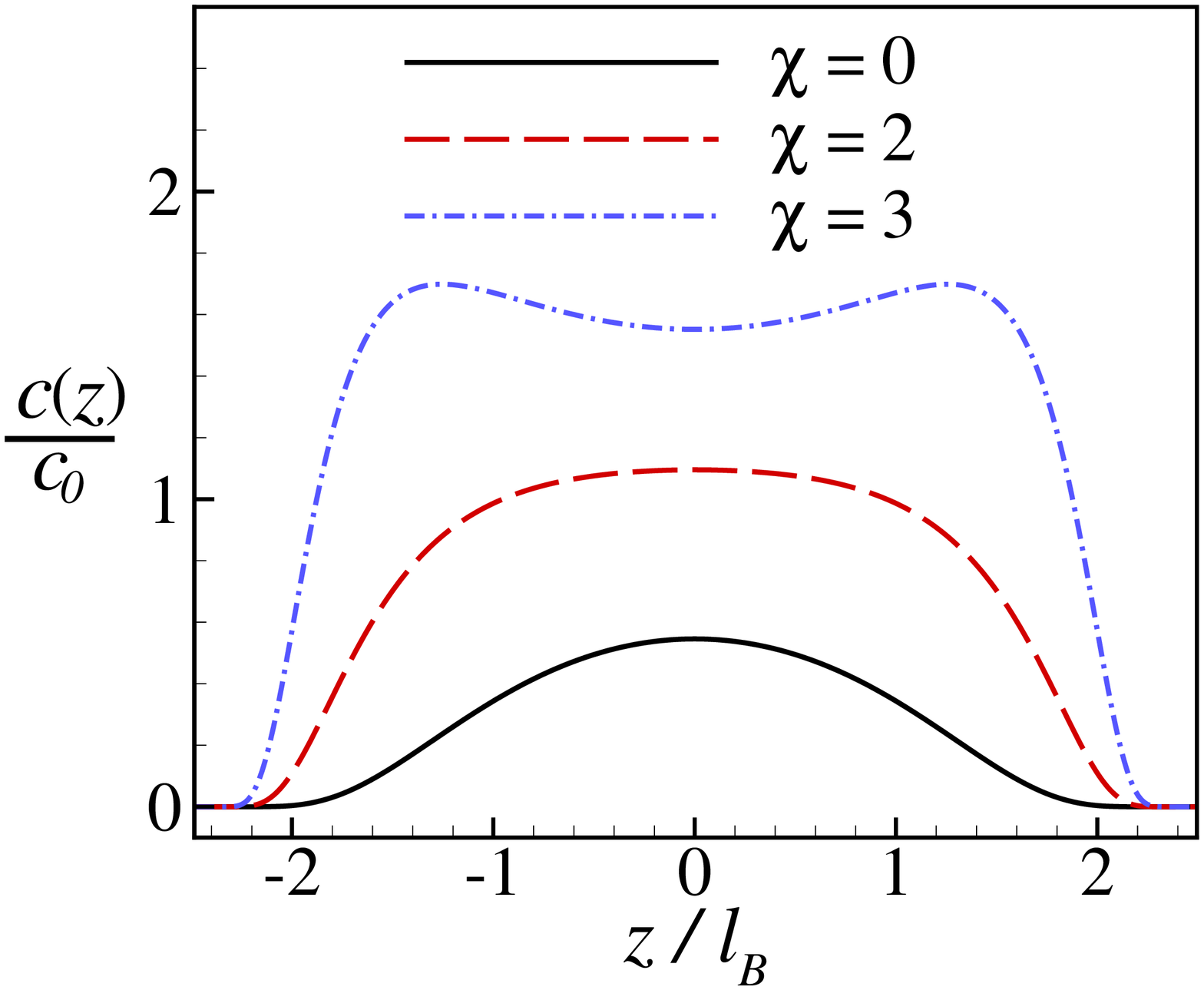} (b)
	\end{center}\end{minipage} \hskip0.1cm	
\begin{minipage}[t]{0.325\textwidth}\begin{center}
		\includegraphics[width=\textwidth]{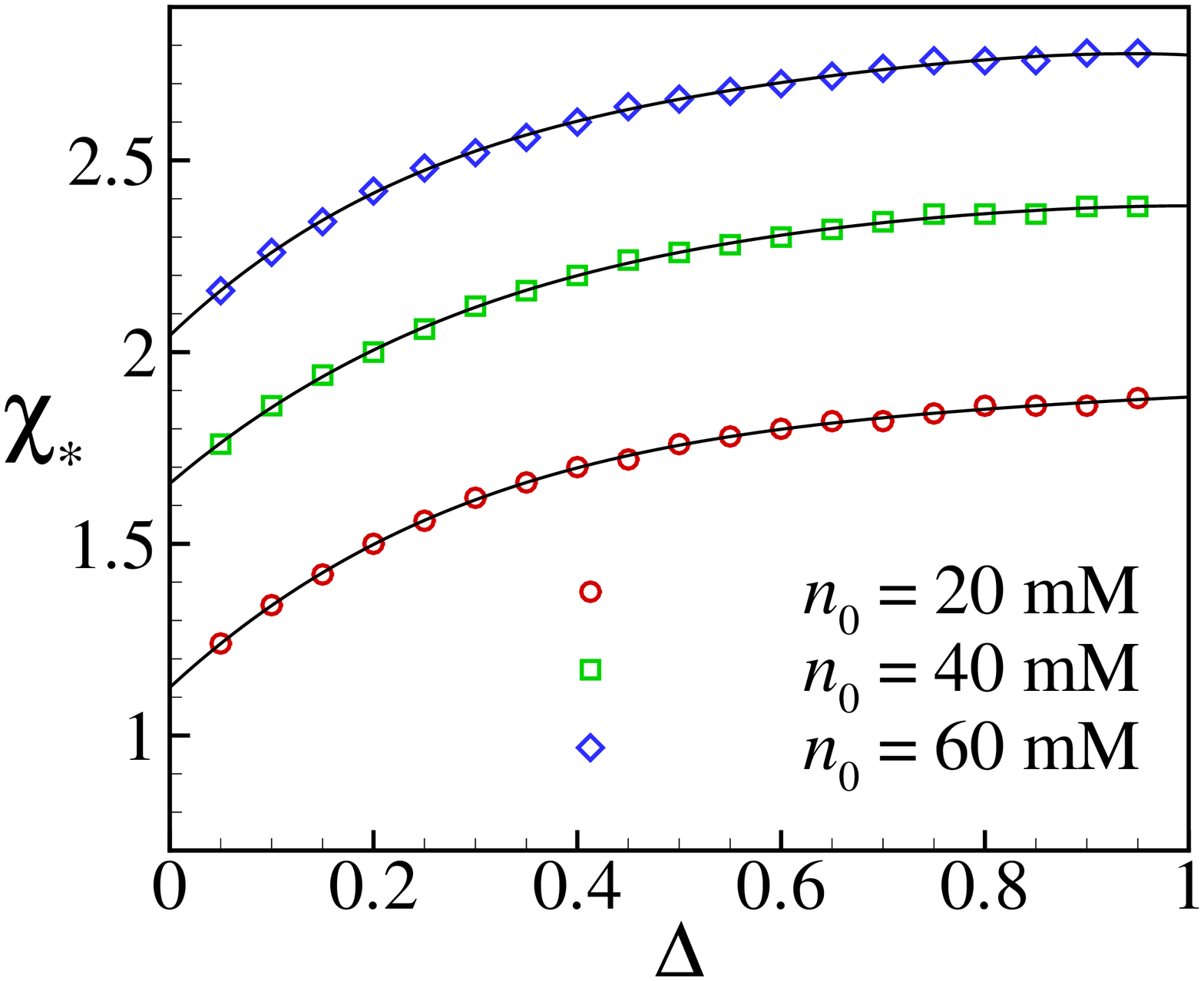} (c)
	\end{center}\end{minipage}
\caption{Rescaled density profile of dressed multivalent ions as a function of the rescaled normal position in the slit between the randomly charged inner surfaces of two net-neutral  dielectric slabs for  (a) dielectrically homogeneous system with $\Delta=0$ and (b) dielectrically inhomogeneous system with $\Delta=0.95$ (appropriate for the water/hydrocarbon interface). Other parameters are fixed as $d/\lB=5$ and $\kappa\lB=0.35$ ($c_0=1$~mM, $n_0=20$~mM). 
Panel (c) shows the threshold value, $\chi_\ast$ (above which the unimodal density profile changes to a bimodal one) as a function of $\Delta$ for $d/\lB=5$, $c_0=1$~mM and $n_0=20, 40$ and 60~mM.}
\label{f:neutral_dist}
\end{center}\end{figure*}

In the two-slab geometry, we can then re-express the (number) density profile of multivalent ions in the slit region, $-d/2 \leq z \leq d/2$, as
\begin{equation}
c(z)= c_0 \rme^{-\beta u (z)},
\end{equation}
where
\begin{eqnarray}
\label{eq:ResU_neutral}
 \nonumber u(z)&=&\frac{q^2 e_0^2}{4\pi} \int_0^\infty Q \rmd Q\, \hat G(Q; z,z)\\
 \nonumber && -\frac{\beta q^2  g e_0^4}{4\pi}\!\int_0^\infty\! \!Q \rmd Q \big[\hat G^2(Q;z, -d/2)+\hat  G^2(Q;z,d/2)\big].\\
\end{eqnarray}
The {\em interaction} free energy of the system (per $k_{\mathrm{B}}T$ and per unit area, $S$) can  be written as
\begin{equation}\label{e:F_neutral}
\frac{\beta \mathcal F}{S}= g \lB f(\kappa, d, \Delta)-\int_{-d/2}^{d/2} \rmd z\, c(z),
\end{equation}
where we have subtracted additive terms that are independent of surface separation $d$ and defined
\begin{equation}\label{e:f}
f(\kappa, d, \Delta)\equiv \int_{0}^{\infty} Q dQ \,\frac{\Delta_s(1+\Delta_s)^2}{\gamma(\rme^{2d\gamma}-\Delta_s^2)}.
\end{equation}

\section{Results}
\label{sec:results}

\subsection{Distribution of multivalent ions\label{ss:Dis}}

Let us first focus on the case of a dielectrically homogeneous system with $\Delta=0$. The density profiles of multivalent ions for this case are shown in Fig. \ref{f:neutral_dist}a, where we have rescaled the density profiles with their bulk value $c_0$ and the position in the slit $-d/2<z<d/2$ with the Bjerrum length $\lB$. The  inter-surface distance and the screening parameter are rescaled in the same way. In the figure, we have fixed $d/\lB=5$ and $\kappa\lB=0.35$, 
equivalent to mono- and multivalent salt concentrations of $n_0=20$~mM and $c_0=1$~mM when the Bjerrum length is taken as $\lB=0.71$~nm (appropriate for water at room temperature, i.e., with $T=293$~K and $\epsilon_m=80$) and the multivalent ion valency is taken as $q=4$ (note that throughout this paper we focus on the case of asymmetric Coulomb fluids with tetravalent ions but the generalization of our results to other values of $q$ is straightforward).

The surface charge disorder variance is shown in the figure in terms of the dimensionless {\em disorder coupling (or strength) parameter} \cite{jcp2014,sm2015}
\begin{equation}
\chi = 2\pi q^2\lB^2 g,
\end{equation}
which is varied in the figure as $\chi=0, 1$ and 2 (corresponding to strictly neutral surfaces with $g=0$, and randomly charged surfaces with $g=0.02$~nm$^{-2}$ and 0.04~nm$^{-2}$, respectively).

As seen, for strictly neutral (or disorder-free) dielectrics, the density profile (black solid curve) shows a partial depletion of multivalent ions from the vicinity of the surface boundaries. This is caused by the salt-image repulsion (corresponding to the first term in Eq. (\ref{eq:u2})), which, as noted in the previous Section,  are caused by the discontinuity in the distribution of salt ions across the dielectric interfaces, giving a non-vanishing $\Delta_s$ according to Eq. (\ref{eq:Delta_s_homogen}). However, when the surfaces are randomly charged (dashed curves), the situation turns out significantly different and the multivalent ions are attracted quite strongly towards the surfaces despite the  ``soft" salt-image repulsions (moreover, one can note from the shown profiles that a larger amount of multivalent ions are pulled into the slit from the bulk solution). In fact, the resulting disorder potential acting on individual multivalent ions (corresponding to the second term in Eq. (\ref{eq:u2})) exhibits a singular (logarithmic) behavior at the two surfaces and gives an algebraically diverging density as $c(z)\sim (d^2/4 - z^2)^{-\chi/2}$ when $z\rightarrow \pm d/2$. This kind of phenomena has been discussed in detail in the context of disordered charged surfaces bearing a net charge density \cite{jcp2014,sm2015}, where we show that the singular behavior of the density profile is closely connected with the {\em anti-fragile} behavior of multivalent counterions: Adding a quenched disorder component to an otherwise uniform distribution of surface charges leads to a lowered entropy (or thermal disorder) for the counterions. In the present context, with net-neutral surfaces, the disorder-induced effects are qualitatively similar as they are regulated only by the variance of the charge distribution, $g$, rather than the net charge density of the surface.

Finite dielectric discontinuity at the bounding surfaces, $\Delta>0$, creates a much stronger depletion effect than in the case of dielectrically homogeneous system (see Fig. \ref{f:neutral_dist}b). In fact, the density profile of multivalent ions vanishes in the immediate vicinity of the surfaces because of the strong dielectric-image repulsion, provided again by the first term in Eq. (\ref{eq:u2}). This term is singular itself and diverges on approach to the surface as $\sim 1/(d/2\mp z)$ when $z\rightarrow \pm d/2$, thus  dominating over the singular potential created by the charge randomness on the boundaries and, leading to the vanishing contact density of multivalent ions regardless of the disorder strength, $\chi$, as seen in the figure (even though a larger amount of multivalent ions are found in the slit at larger values of $\chi$). The competition between these two mechanisms of repulsion and attraction leads to a change in the shape of the density profile from unimodal to bimodal beyond a threshold value of the disorder strength, $\chi_*$. The data in Fig. \ref{f:neutral_dist}c show $\chi_*$ as a function of the dielectric discontinuity parameter, $\Delta$, for fixed $d/\lB=5$, $c_0=1$~mM and different values of  $n_0=20, 40$ and 60~mM (corresponding to $\kappa\lB=0.35, 0.48$ and 0.58, respectively). Clearly, for larger salt screening and/or dielectric discontinuity parameter,  a larger degree of charge disorder is required in order to counteract the image-charge repulsions. These features of the density profiles are qualitatively similar to those found in the case of charged surfaces \cite{jcp2014,sm2015} and, therefore, we shall not delve further into the details of the behavior of the density profile and proceed with the analysis of the effective interaction between net-neutral surfaces.

\subsection{Effective interactions}

Effective interactions between neutral dielectric slabs are standardly described in terms of the vdW interactions as, for instance, formulated within the Lifshitz theory \cite{Parsegian2005,Ninham76}.
The vdW interaction pressure for two plane-parallel slabs can be  expressed as \cite{Parsegian2005}
\begin{equation}
\label{p:vdwz}
P_{vdW}= - k_{\mathrm{B}}T\int_0^\infty \frac{Q \rmd Q}{2\pi }\frac{\gamma \Delta_s^2 \,\rme^{-2 \gamma d}}{1 - \Delta_s^2\, \rme^{-2 \gamma d}} -\frac{A }{6\pi d^3},
\end{equation}
where the first term comes from the thermal zero-frequency mode of the  electromagnetic field-fluctuations and the second term comes from the higher-order Matsubara frequencies, pertaining to quantum fluctuations. 
$A$ is then the quantum part of the Hamaker coefficient, which, based on an upper bound estimate in the case of hydrocarbon slabs interacting across an aqueous medium, can be taken as $A=3\, \text{zJ}$ \cite{hamakar,SCdressed3}.  
The vdW interaction is dominant mostly at the scale of a few nanometers for the inter-surface distance \cite{French-RMP}.

In the present model, one needs to account for the electrostatic contribution to the inter-surface pressure, $P_{es}$, as well. This contribution can be written as the difference between the slit pressure of multivalent and monovalent ions and the bulk pressure $P_b =(n_b+c_0) k_{\mathrm{B}}T$, where $n_b = 2n_0+qc_0$ is the total bulk concentration of monovalent ions, i.e., $P_{es}=P_s-P_b$. The slit pressure can be calculated as $P_s = -{\partial {\mathcal F}}/({S \partial d})+n(0)k_{\mathrm{B}}T$, in which the first term is the contribution of multivalent ions that follows from the derivative of the free energy expression, Eq. (\ref{e:F_neutral}),  w.r.t. the inter-surface distance when all other parameters are kept constant, while the second term is the contribution of monovalent ions that has been expressed in terms of the total mid-plane density of monovalent ions \cite{SCdressed3}. This latter quantity can be estimated here through the relation $n(z)=n_b \exp[-\beta u(z)]\big|_{q=1}$ for $z=0$, which has been shown by means of explicit Monte-Carlo simulations of a disorder-free system \cite{perspective,SCdressed3} to give an accurate estimate of the distribution of monovalent ions in the slit.

The electrostatic pressure can thus be decomposed into its different components as
\begin{equation}
P_{es}=P_{dis}+ P_c+ P_{mon},
\label{eq:P_total}
\end{equation}
where the components are obtained explicitly as
\begin{eqnarray}
\label{eq:P_dis}
 && \beta P_{dis}=-g \lB\frac{\partial f(\kappa, d, \Delta)}{\partial d}\\
\label{eq:P_c}
  && \beta P_c=\frac{\partial}{\partial d}\left[\int_{-d/2}^{d/2} c(z) dz\right] -c_0\\
\label{eq:P_mon}
 && \beta P_{mon}=n(0)-n_b.
\end{eqnarray}
The contribution $P_{dis}$ arises from self-interactions of random charges on the  surfaces of the two slabs with their (dielectric and/or salt) image charges. This contribution stems from the first term of the interaction free energy (\ref{e:F_neutral}) and is present irrespective of multivalent ions, being typically comparable in range and strength with the vdW pressure. It can, however, scale differently with the inter-surface distance and can be repulsive or attractive depending on the sign of dielectric discontinuity parameter, $\Delta$ (e.g., it is repulsive for $\Delta>0$, applicable to the cases studied in this paper), resulting thus in a rather diverse behavior for the inter-surface interaction between two net-neutral slabs in the absence of multivalent ions \cite{rudiali,pre2011}. The two terms $P_c$ and $P_{mon}$ correspond to the disjoining or osmotic pressure contributions of multivalent and monovalent ions, respectively. The former stems from the second term of the interaction free energy (\ref{e:F_neutral}), while, as noted above, the latter is included heuristically through Eq. (\ref{eq:P_mon}).

\begin{figure}[t!]
 \begin{minipage}[h]{0.4\textwidth}\begin{center}
		\includegraphics[width=\textwidth]{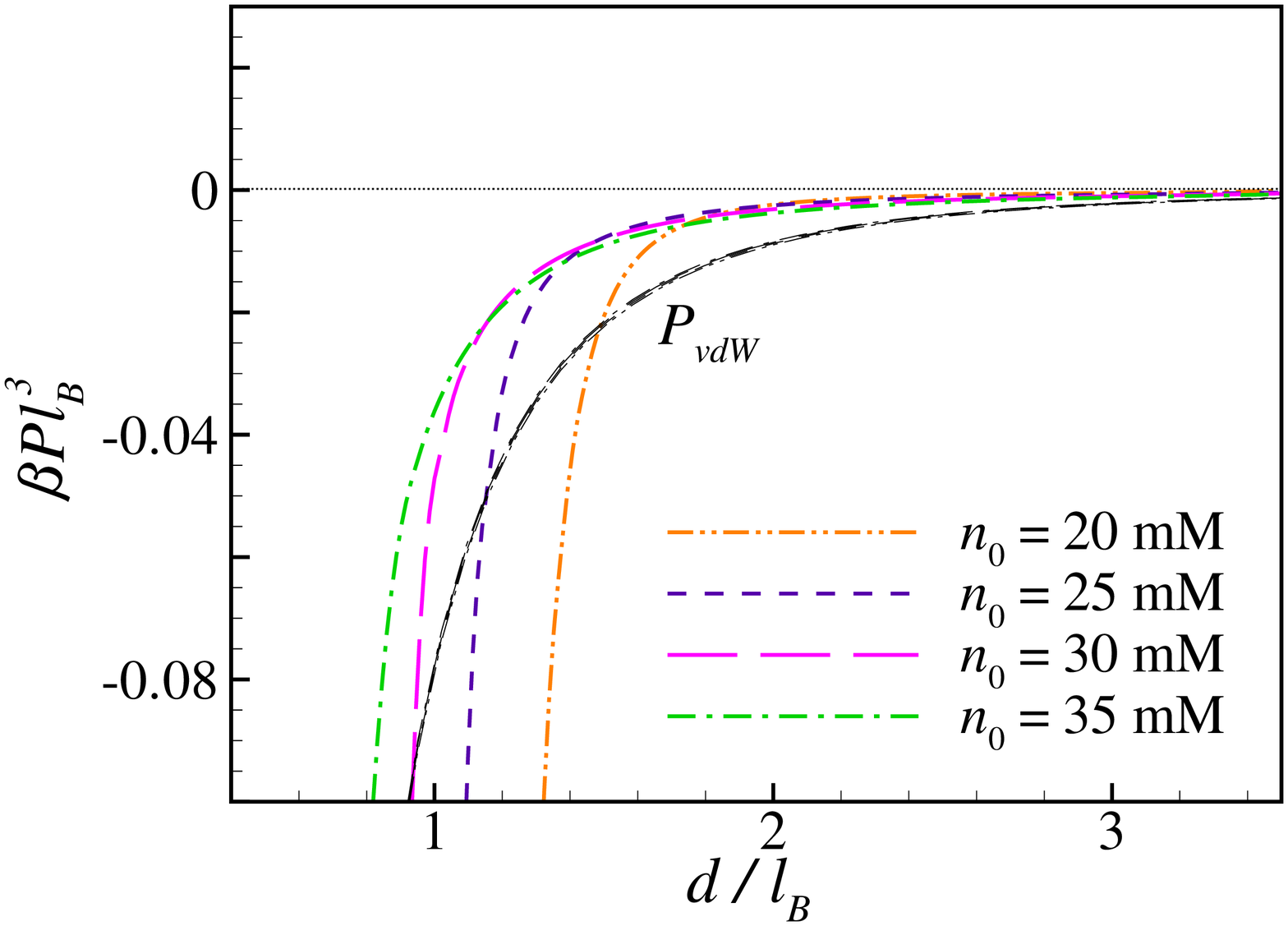}(a)
	\end{center}\end{minipage}\vskip 0.3cm	
    \begin{minipage}[h]{0.4\textwidth}\begin{center}
		\includegraphics[width=\textwidth]{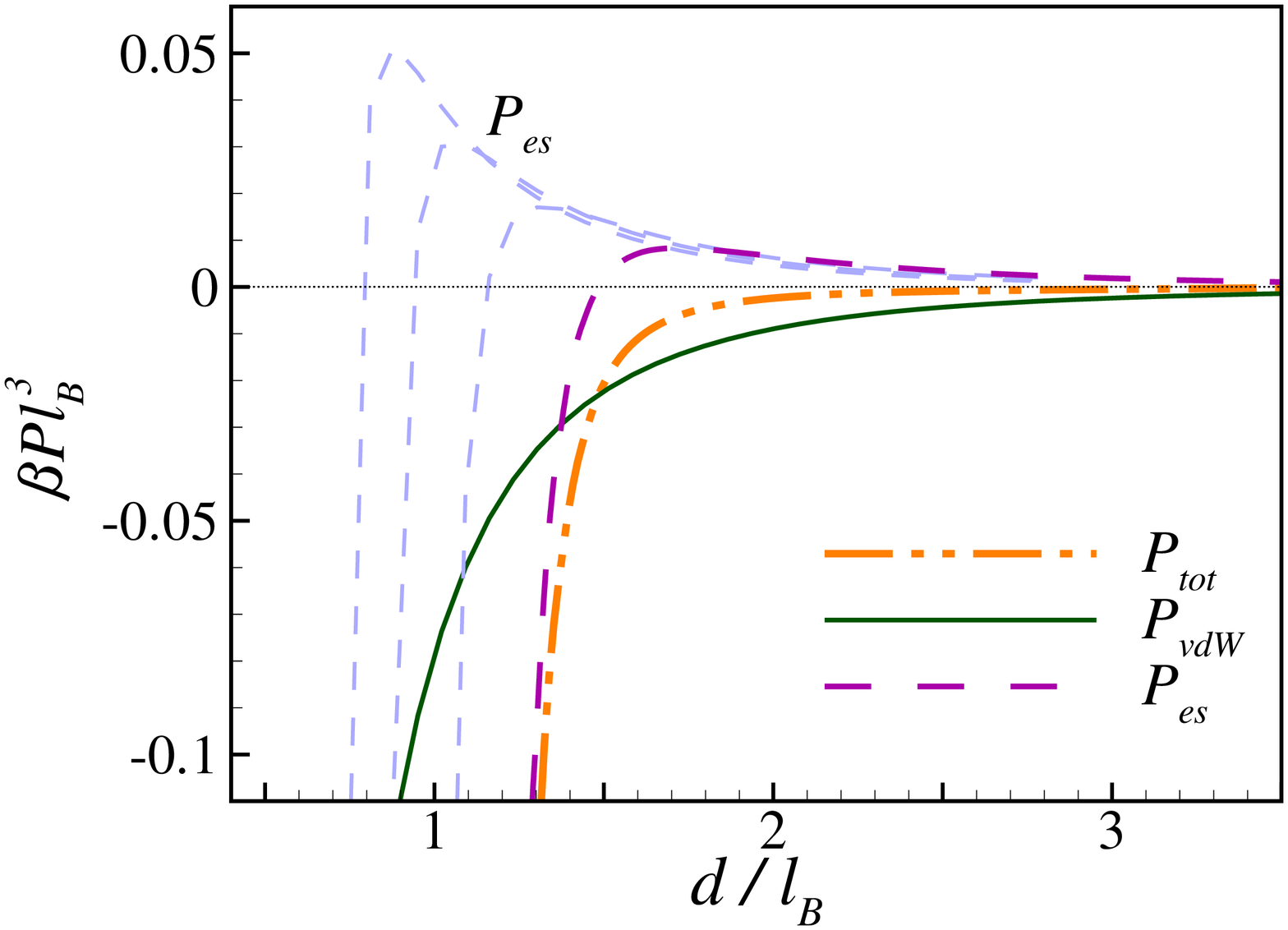}(b)
	\end{center}\end{minipage}
 \caption{(a) Rescaled total interaction pressure as a function of the rescaled distance between the randomly charged inner surfaces of two net-neutral dielectric slabs for fixed $\chi=2$, $c_0=1$~mM and $n_0=20, 25, 30$ and 35~mM as indicated on the graph (colored dashed curves). The corresponding vdW pressure for these parameter values closely overlap (black dashed curves). Panel (b) shows the total pressure ($P_{tot}$) along with its electrostatic ($P_{es}$) and vdW ($P_{vdW}$) components for $c_0=1$~mM and $n_0=20$~mM. The light-colored dashed curves show $P_{es}$ for $n_0=25, 30$ and 35~mM (from right to left). 
 }
  \label{f:neut_kappa}
\end{figure}

\begin{figure*}[t]\begin{center}
	\begin{minipage}[t]{0.33\textwidth}\begin{center}
       (a) \includegraphics[width=\textwidth]{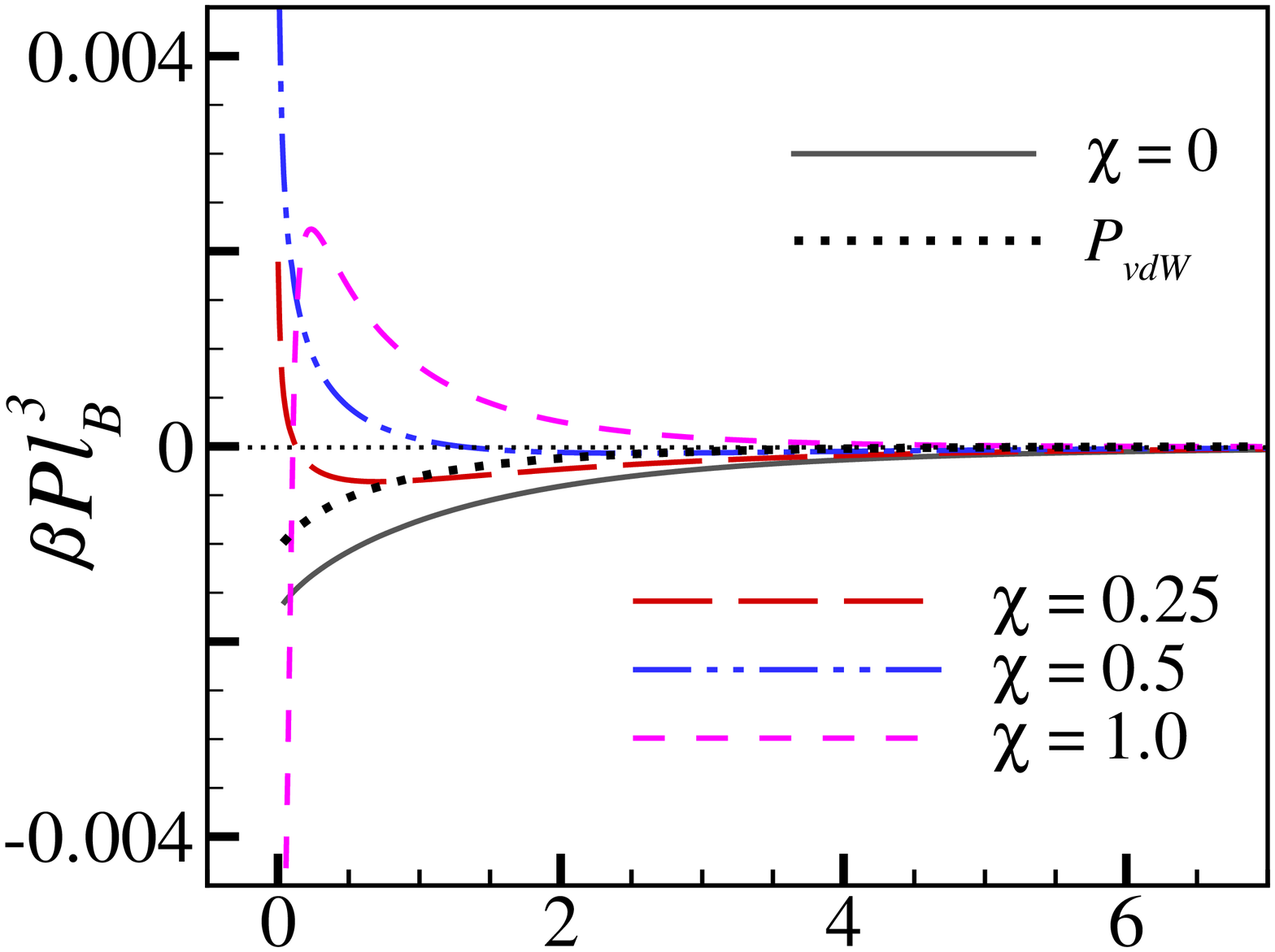} \\
       \vskip-0.4cm 
        \includegraphics[width=\textwidth]{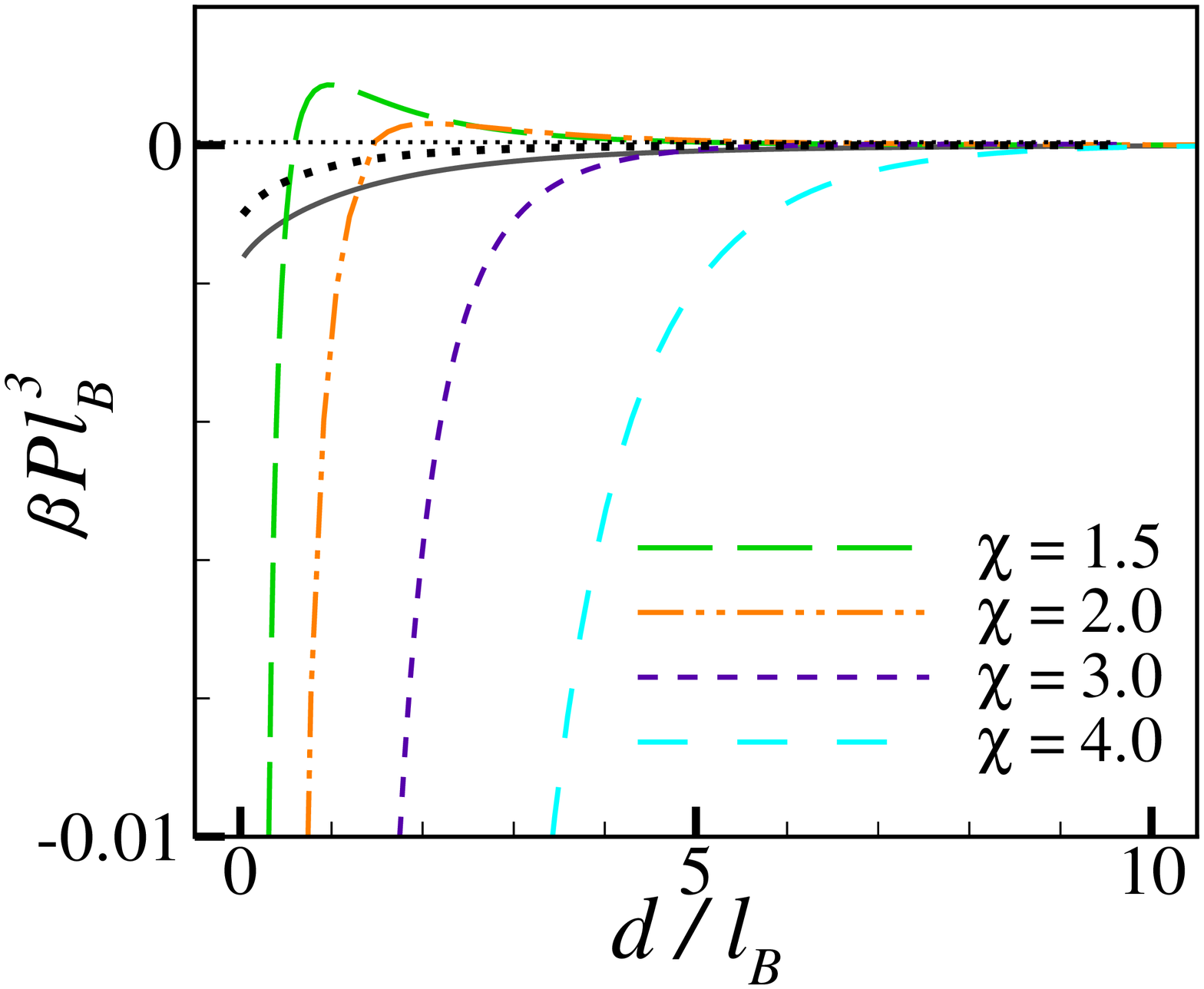} (b)
	\end{center}\end{minipage}
	\begin{minipage}[t]{0.33\textwidth}\begin{center}
		(c) \includegraphics[width=\textwidth]{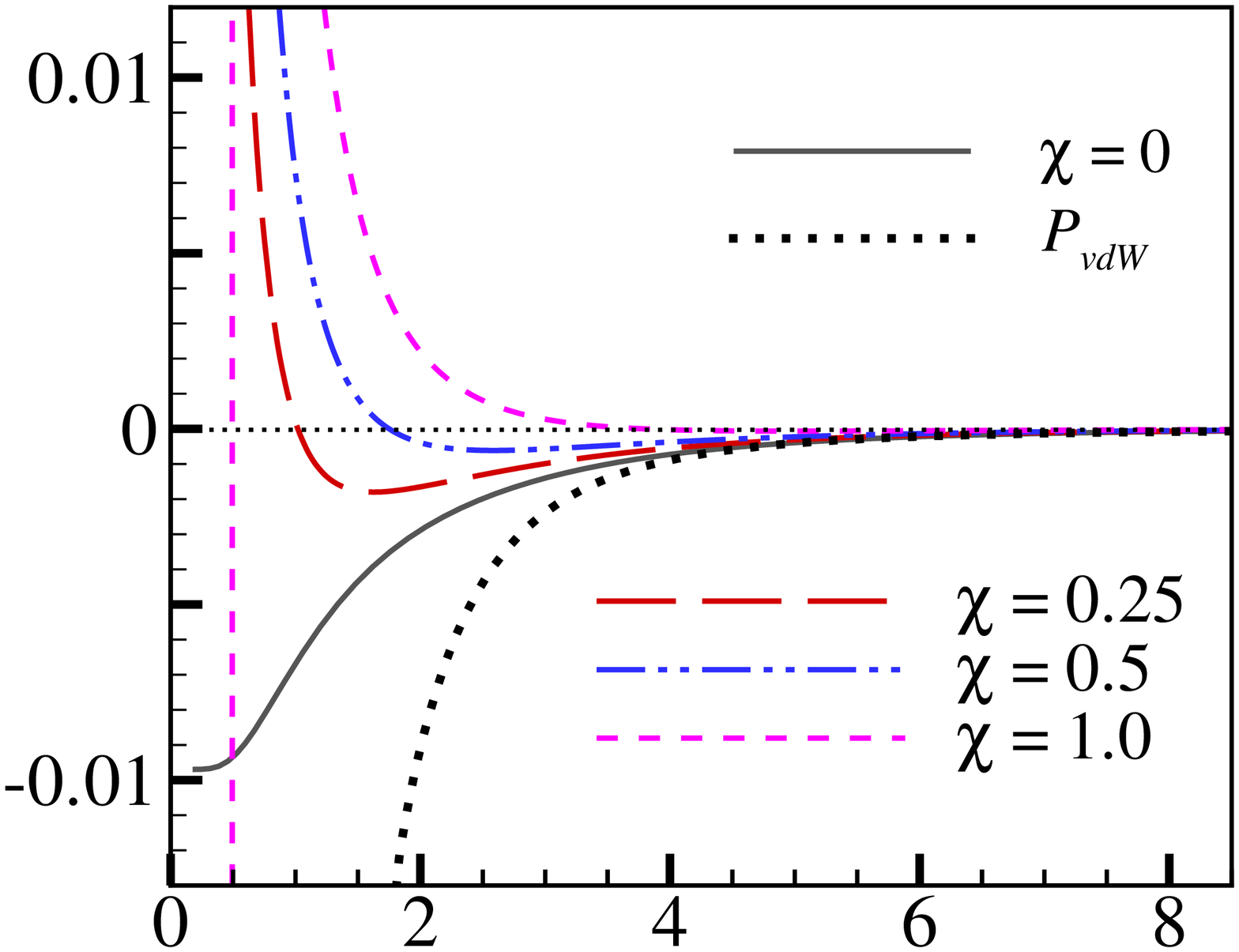} \\
        \vskip-0.4cm 
      \includegraphics[width=\textwidth]{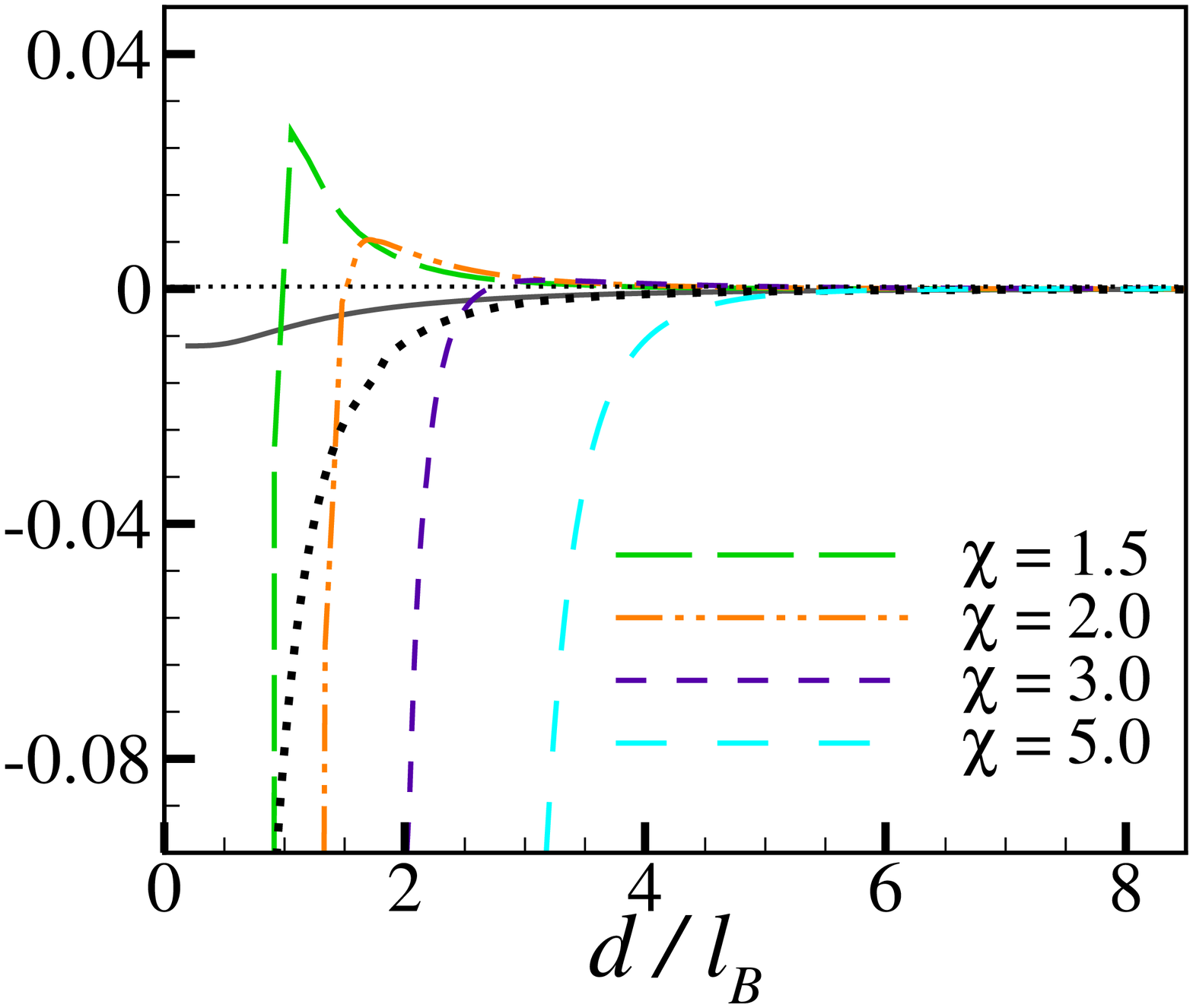} (d)
	\end{center}\end{minipage} 
	\begin{minipage}[t]{0.33\textwidth}\begin{center}
		(e) \includegraphics[width=\textwidth]{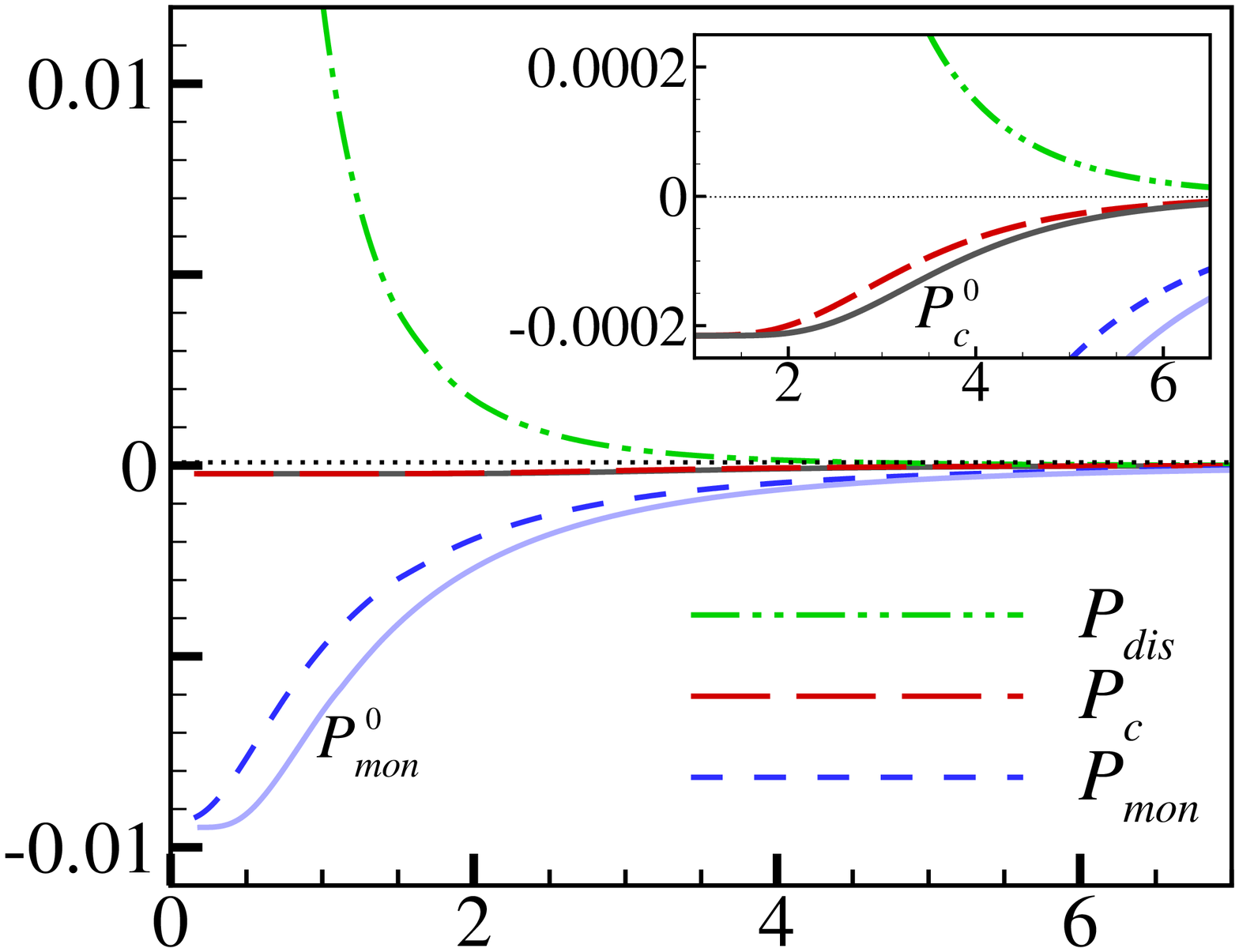} \\
       \vskip-0.4cm 
       \includegraphics[width=\textwidth]{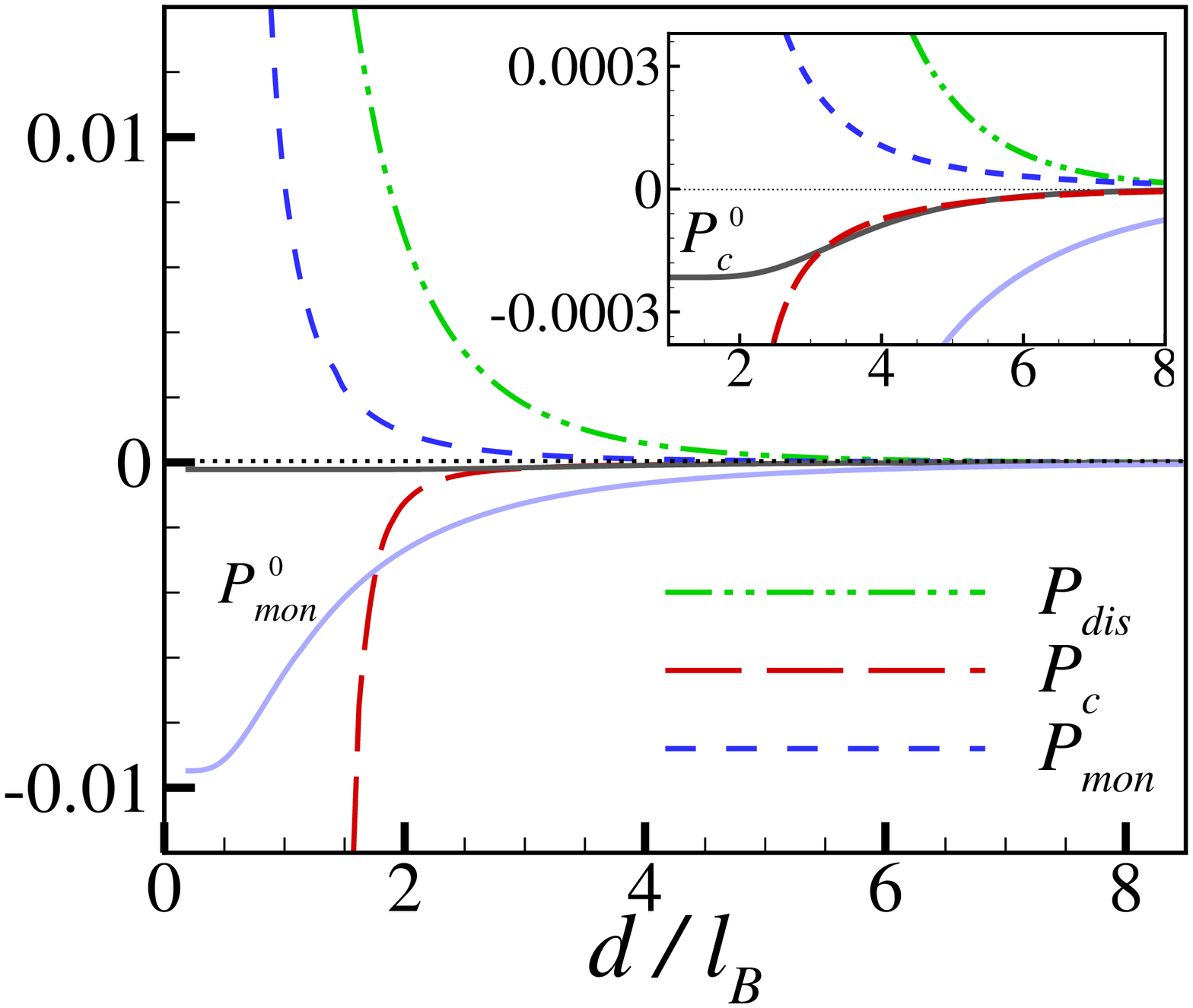} (f)	
       \end{center}\end{minipage} 
\caption{Rescaled electrostatic pressure as a function of the rescaled distance between the randomly charged inner surfaces of two net-neutral dielectric slabs for fixed $k\lB=0.35$ ($c_0=1$~mM and $n_0=20$~mM. Panels (a) and (b) show the results for weakly and strongly disordered cases in a dielectrically homogeneous system ($\Delta=0$), while panels (c) and (d) show the same for $\Delta=0.95$. The results for the disorder-free case (black solid curve) and the vdW pressure (black dotted curves) are shown for comparison. Panels (e) and (f) show the contributing components to the electrostatic pressure, $P_{es}$, in the examples of a weakly disordered system with $\chi$=0.5 and a strongly disordered system with $\chi$=2, respectively.  The other parameters are as in Panels (c) and (d). For comparison, we also show the contribution of mono- and multivalent ions in the disorder-free case, $P_{mon}^0$ and $P_c^0$. Insets show the rescaled pressure components over a smaller range of  values.}
\label{f:neutral_pressure}
\end{center}\end{figure*}

The total pressure acting on the slabs is finally given by
\begin{equation}
P=P_{es}+P_{vdW}.
\end{equation}
This quantity is plotted in Fig. \ref{f:neut_kappa}a as a function of the rescaled inter-surface distance for fixed $\chi=2$,  $c_0=1$~mM and four different values of $n_0=20, 25, 30$ and 35~mM, which correspond to $ \kappa\lB=0.35, 0.38, 0.42$ and 0.45, respectively  (colored dashed curves).
For these same parameter values, we also show the vdW pressure in Fig. \ref{f:neut_kappa}a (black dashed curves), but these latter curves closely overlap and are not discernible at the implied resolution. This clearly indicates that the dependence of the total pressure on the salt screening parameter enters mainly through the electrostatic contribution (see below). In fact, as one can see from the figure, the total pressure, $P$, is less attractive than the standard vdW pressure at large separations, but, as $n_0$ decreases, the total pressure becomes more strongly attractive as compared with the vdW contribution. We show the electrostatic and vdW components of the total pressure separately in Fig. \ref{f:neut_kappa}b for $c_0=1$~mM and $n_0=20$~mM ($ \kappa\lB=0.35$). The electrostatic contribution, $P_{es}$, shows a non-monotonic behavior: It is repulsive and decays to zero at large separations, while it becomes strongly attractive  and diverges at small separations, showing thus a weakly repulsive maximum at intermediate inter-surface separations. The light-colored dashed curves show the same quantity for larger values of  $n_0=25, 30$ and 35~mM 
(from right to left). Clearly, the extent of the repulsion and the non-monotonic behavior of the electrostatic pressure become more pronounced as  $n_0$ is increased. Non-monotonic interaction pressures due to  image-induced ion depletion have also been found in the absence of surface charge disorder but at relatively large bulk salt concentrations (e.g., above 250mM) \cite{Kjellander08}.

Before proceeding with a more detailed analysis of the electrostatic contribution, we emphasize here that $P_{es}$ and $P_{vdW}$ show clearly different qualitative behaviors as one can see from Fig. \ref{f:neut_kappa}b. The standard DLVO description of colloidal interactions \cite{Israelachvili,VO} in terms of the vdW interaction of neutral surfaces is therefore insufficient when the surfaces are neutral only {\em on the average} and otherwise carry random positive and negative charges, and/or when the system contains also an asymmetric Coulomb fluid. Our results show that only a small degree of charge randomness with, e.g., a surface charge variance of $g=0.04$~nm$^{-2}$ (corresponding to $\chi=2$), will be enough to generate a sizable deviation from the standard vdW prediction.

\subsection{Interplay between charge disorder and multivalent ions}

In order to gain further insight into the intriguing role of disorder-induced effects in the presence of multivalent ions, we examine the behavior of the effective electrostatic pressure, $P_{es}$, and its components in more detail.

First, we consider two strictly neutral (disorder-free) surfaces, in which case the disorder effects vanish and the electrostatic inter-surface pressure, $P_{es}$, is given only by  the osmotic pressures of multivalent and monovalent ions. The results are shown by black solid curves in Figs.~\ref{f:neutral_pressure}(a, b) and Figs.~\ref{f:neutral_pressure}(c, d)  for a dielectrically homogeneous ($\Delta=0$) and a dielectrically inhomogeneous system ($\Delta=0.95$), respectively. For the latter case, the contributing components of the electrostatic pressure, i.e., $P_c^0$ and $P_{mon}^0$, are plotted in Figs.~\ref{f:neutral_pressure}(e, f) for comparison. The vdW pressure is shown by black dotted curves in panels a to d. In all cases, we have taken $c_0=1$~mM and $n_0=20$~mM.

It turns out that the electrostatic pressure in a disorder-free system ($\chi=0$), which is given by $P_{es}=P_{mon}^0+P_c^0$,  is negative (attractive)  for all inter-surface separations and tends to zero at large separations because of salt screening effects. This means that the negative bulk pressure is stronger in magnitude than the slit pressure for both cases of $\Delta=0$ and 0.95. This is because of the image-induced depletion of ions generated by salt and/or dielectric-image charges that are of the same sign when the surfaces are immersed in a medium of larger dielectric constant as assumed  here. In the limit $d\rightarrow0$, $P_{es}$ in an inhomogeneous system reduces to the bulk pressure, $P_b =(n_b+c_0) k_{\mathrm{B}}T$, due to complete depletion of ions from the slit under the strong repulsive force of dielectric-image charges, while in a dielectrically homogeneous system, complete depletion is not achieved; hence, $P_{es}$ is less attractive in this latter case. On the other hand, for the given parameter values, the vdW pressure is more (less) attractive than the electrostatic pressure, $P_{es}$, for $\Delta=0.95$ ($\Delta=0$) as one can see from the figures.

Next, we discuss how the introduction of charge disorder on the inner surfaces of the net-neutral dielectrics affects the effective electrostatic interaction between them. To this end, we vary the disorder strength parameter from $\chi=0$ up to $\chi=5$ (corresponding to $g=0-0.1$~nm$^{-2}$) and divide the results into two categories of ``weakly'' and ``strongly'' disordered cases.

In the {\em weak disorder regime} (typically $\chi\lesssim 1$), the electrostatic interaction pressure, $P_{es}$, becomes gradullay more repulsive or positive (especially at small separations) as compared to that of strictly neutral surfaces when $\chi$ is increased up $\chi=1$ (see colored dashed curves in Figs.~\ref{f:neutral_pressure}a and c for $\Delta=0$ and $\Delta=0.95$, respectively). The repulsive interaction pressure is stronger for larger  values of the dielectric discontinuity parameter, $\Delta$ (compare panels a and c). The most remarkable feature of our results is that $P_{es}$ in this regime adopts a non-monotonic behavior: For very small $\chi$, it develops a point of zero electrostatic pressure followed by a shallow attractive minimum  at intermediate separations  (see, e.g., $\chi=0.25$, red dashed curve, in panel a, or $\chi=0.25$ and 0.5, red and blue dashed curves in panel c). The depth of this minimum decreases and the interaction profile instead develops a repulsive (positive) hump at large values of $\chi$ (e.g., $\chi=1$). These behaviors are in clear contrast to what is observed in the case of strictly neutral (disorder-free) surfaces (black solid curves).

In order to elucidate the origin of this difference, we compare different components of $P_{es}$ in the case of an inhomogeneous system with $\Delta=0.95$ and $\chi=0.5$  in Fig.~\ref{f:neutral_pressure}e.  For comparison, the light-blue solid curve and the black solid curve (which is seen more clearly in the inset) show the osmotic contribution of monovalent and multivalent ions, $P_{mon}^0$ and $P_c^0$,  in the case of strictly neutral surfaces ($\chi=0$). Both contributions are attractive (negative), while the former is the dominant one as expected since we have assumed that  multivalent ions enter in small bulk concentrations (here, $c_0=1$~mM). By introducing a weak charge randomness  ($\chi=0.5$ for the relevant curves in Fig.~\ref{f:neutral_pressure}e), the negative pressure of both multivalent ($P_c$, red dashed curves, inset) and monovalent ions ($P_{mon}$, blue dashed curve, main set) decrease in magnitude and, thus, become less attractive as compared with the disorder-free case. It is important to note that the distribution of both monovalent and multivalent ions and, thus, their respective osmotic pressure, $P_{mon}$ and $P_c$, are affected by the surface charge disorder through the disorder-induced, single-ion potential, which is the second term in Eqs. (\ref{eq:u2}) or (\ref{eq:ResU_neutral}). The decrease in the magnitude of attractive pressure components $P_{mon}$ and $P_c$ in the weak disorder regime is   because more mono- and multivalent ions are attracted into the slit from the bulk solution in the presence of surface charge disorder (see Fig.~\ref{f:neutral_dist} and Fig. \ref{f:sum_multi} below). Nevertheless, at very small separations, all ions are again totally depleted due to the repulsive forces of image charges and $P_{mon}$ and $P_c$ reduce to the same bulk values as $P_{mon}^0$ and $P_c^0$ do when $d\rightarrow 0$, i.e., $-n_bk_{\mathrm{B}}T$ and $-c_0 k_{\mathrm{B}}T$, respectively. The crucial role, however, is played by the repulsive disorder self-interaction component, $P_{dis}$ (green dashed curve), whose effect is amplified by increasing the charge disorder strength and/or the interfacial dielectric mismatch as follows from Eq. \eqref{eq:P_dis}. It is the interplay between this repulsive disorder-induced pressure (which dominates at small to intermediate separations) and the attractive osmotic pressure of ions (which dominate at larger separations) that gives rise to the non-monotonic behavior mentioned above. 
For sufficiently small $\chi$, the pressure of multivalent ions is typically too small (as compared to the other two components) to change this qualitative behavior.

This picture changes drastically in the {\em strong disorder regime} (typically $\chi\gtrsim 1$), where the pressure due to multivalent ions becomes a key factor. In  Fig.~\ref{f:neutral_pressure}b and d, we plot $P_{es}$ for larger disorder strength parameters, $\chi=1.5$ up to 5 for $\Delta=0$ and $\Delta=0.95$, respectively. In this regime, we find a reverse trend caused by the surface charge disorder: The repulsive hump at intermediate separations now diminishes and  eventually disappears when $\chi$ is increased to larger values; one thus finds a highly attractive (negative) interaction pressure with a range that can be much larger than that of the vdW interaction pressure (black dotted curves), or the mere image-induced, ion-depletion pressure (black solid curves with $\chi=0$), at sufficiently large disorder strengths.

The different components contributing to the electrostatic pressure for this case are shown in Fig.~\ref{f:neutral_pressure}f for $\chi=2$ with other parameters being the same as Figs.~\ref{f:neutral_pressure}c and d. As seen, while the disorder self-interaction contribution, $P_{dis}$, has become only slightly more repulsive (green dashed curve), the other two components, $P_c$ and $P_{mon}$, show significant changes as compared to their counterparts in the weak disorder regime in Fig.~\ref{f:neutral_pressure}e. Specifically, the attractive contribution due to the osmotic pressure of multivalent ions  (red dashed curve) becomes much stronger than what we find in the weakly disordered or disorder-free cases, while the contribution from monovalent ions now  becomes repulsive (positive) in contrast to its attractive (negative) behavior in these latter cases  (compare panels e and f). The monovalent contribution becomes positive because of an increase in the number of monovalent ions in the slit that creates a larger osmotic pressure. The average number of multivalent ions also increases in the slit as $\chi$ increases; however, these ions correlate strongly with the disorder charges on the bounding surfaces of the slabs (through the second term in Eq. (\ref{eq:ResU_neutral}) which contributes directly to $P_c$, Eq. (\ref{eq:P_c})) and, as a result, give rise to an effective attraction between the net-neutral slabs through the pressure component $P_c$. This effect is in clear contrast with the image-induced, ion-depletion mechanism, in which a {\em decrease} in the number of mono- and multivalent ions in the slit gives rise to attractive osmotic pressures on the slabs similarly from both types of ions. This behavior originates from the (singular) attractive, single-ion potential, which is created by surface charge disorder. It is thus one of the most fundamental aspects of the coupling between surface charge disorder and mobile multivalent ions that follows from our results. 

\begin{figure}[t!]
\centering
   \begin{minipage}[h]{0.4\textwidth}\begin{center}
		\includegraphics[width=\textwidth]{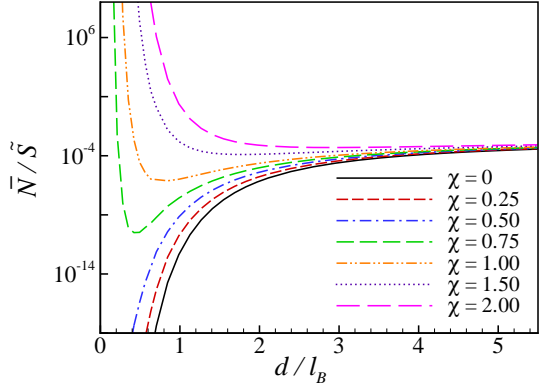}
	\end{center}\end{minipage}	
  \caption{Average total number of multivalent ions in the slit, $\bar N$, per rescaled surface area, $\tilde S=S/\lB^2$, is plotted versus the rescaled inter-surface separation, $d/\lB$, for a system similar to those considered in Figs.~\ref{f:neutral_pressure}c and d. }
  \label{f:sum_multi}
\end{figure}

\begin{figure}[t!]
 \begin{minipage}[h]{0.4\textwidth}\begin{center}
		\includegraphics[width=\textwidth]{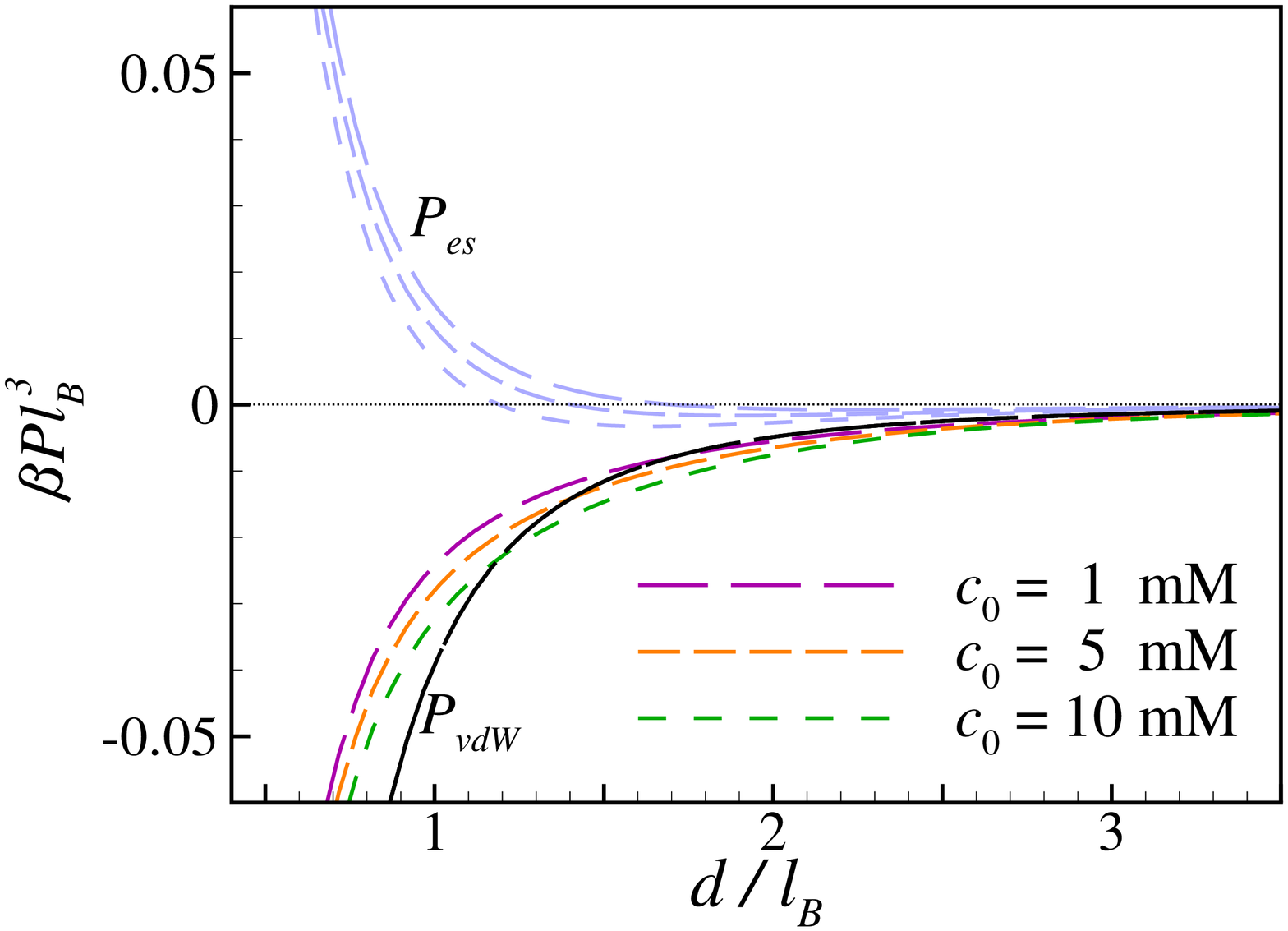}(a)
	\end{center}\end{minipage}\vskip 0.3cm	
    \begin{minipage}[h]{0.4\textwidth}\begin{center}
		\includegraphics[width=\textwidth]{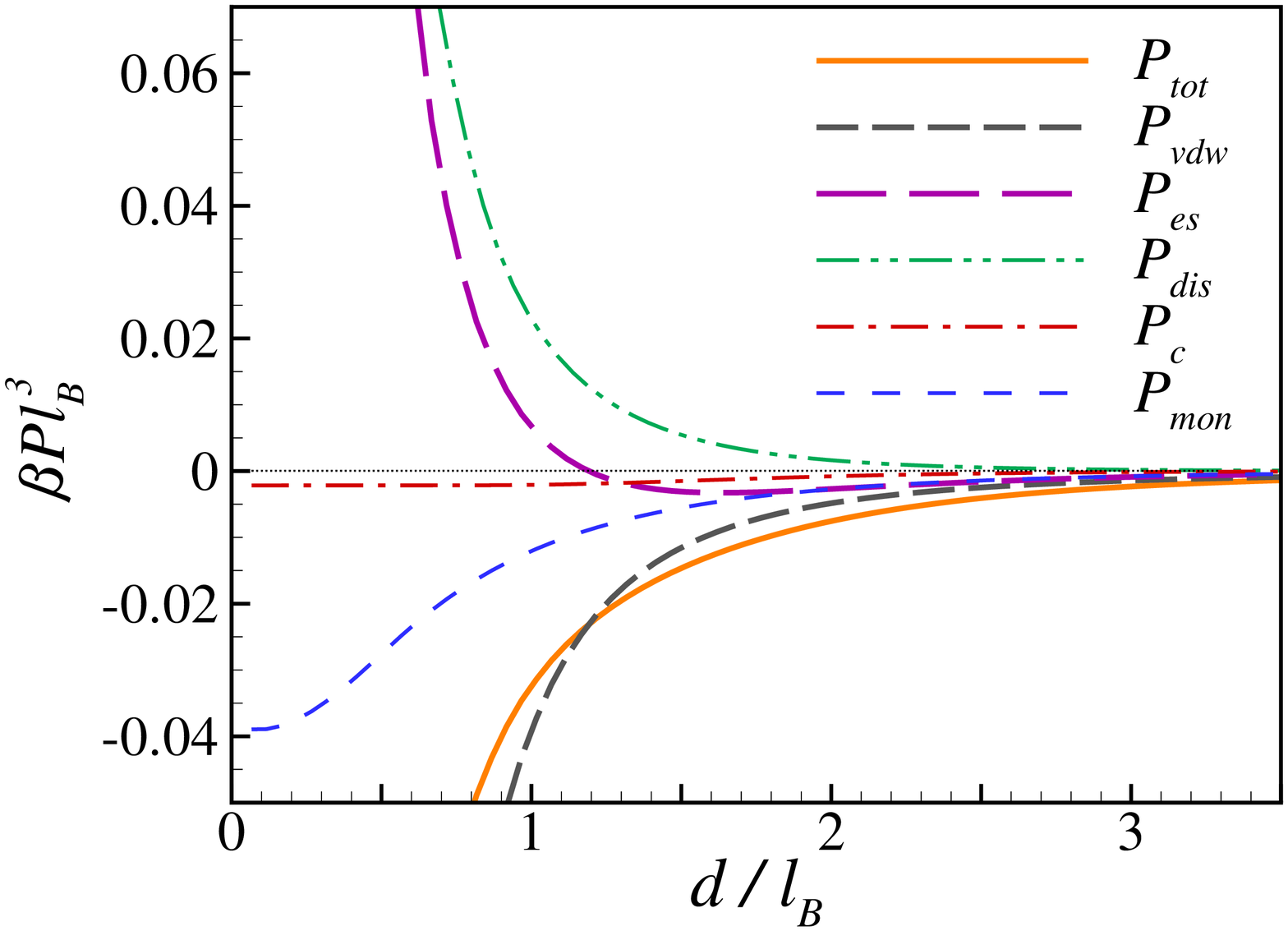}(b)
	\end{center}\end{minipage}
 \caption{(a) Rescaled total inter-surface pressure as a function of the rescaled distance between the randomly charged inner surfaces of two net-neutral dielectric slabs for  fixed $\Delta=0.95$, $\chi=2$, $n_0=100$~mM and $c_0=1$, 5 and 10~mM as indicated on the graph (from top to bottom). The corresponding electrostatic and vdW components of the total pressure, i.e., $P_{es}$ and $P_{vdW}$, are shown in rescaled units as well ($P_{es}$ is shown by the light-colored dashed curves and $P_{vdW}$ by the black one, which closely overlap). Panel (b) shows $P_{tot}$ along with $P_{vdW}$, $P_{es}$ and the three components contributing to the latter, i.e., $P_{dis}$,  $P_c$ and $P_{mon}$  for fixed $c_0=10$~mM and other parameter values as in (a).
 }
  \label{f:kappa_c0}
\end{figure}

The remarks in the preceding discussions on the attraction (depletion) of multivalent ions to (from) the slit region can be corroborated by considering the average total number of ions in the slit. For multivalent ions, this quantity can be calculated from  $\bar N = S\int_{-d/2}^{d/2}\rmd z\, c(z)$. It shows a very different behavior as a function of the inter-surface separation in the weak disorder and strong disorder regimes as one can see in Fig.~\ref{f:sum_multi} (we have used the same parameters as in Figs.~\ref{f:neutral_pressure}c and d). Although, at fixed inter-surface distance, always more multivalent ions are attracted to the slit by increasing the disorder strength, the response of multivalent ions to the decrease in slab separation is quite different for different disorder regimes: For disorder-free and weakly disordered surfaces, multivalent counterions are quickly depleted from the slit by decreasing the slab separations, while, for sufficiently large disorder strengths, these ions (as well as monovalent ions that are not shown here) are more strongly attracted to the slit from the bulk solution due to stronger attraction experienced from the randomly charged, inner surfaces of the slabs. In the intermediate regime of disorder strengths, one find a non-monotonic behavior upon deceasing the inter-surface separation, with the average total number of multivalent ions first decreasing and then increasing at very small separations. One should, however, note that in dielectrically inhomogeneous systems such as the one considered in Fig.~\ref{f:sum_multi}  with $\Delta=0.95$, dielectric-image repulsions eventually win and, thus, even the curves for highly disordered systems  eventually turn downward and we find $\bar N\rightarrow 0$ for $d\rightarrow 0$ or, equivalently, $P_{es}\rightarrow -(n_b+c_0)k_{\mathrm{B}}T$ for all cases in Figs.~\ref{f:neutral_pressure}c and d (these limiting behaviors occur at very small separations that are not physically meaningful, e.g., below $d\simeq 1$~\AA, and, for the sake of presentation, they have not been shown in the plots).


The crossover from weak to strong disorder regimes can be effected also by decreasing the salt screening parameter through a decrease in the bulk concentration of monovalent ions, $n_0$ (and/or increasing the bulk concentration of multivalent ions, $c_0$). For instance, at sufficiently large $n_0$ such as $n_0=100$~mM, the electrostatic interaction pressure, $P_{es}$, at the disorder strength value of $\chi=2$ and multivalent salt concentration of $c_0=1$~mM  shows a weak-disorder behavior (see Fig. \ref{f:kappa_c0}a), in contrast to what we found for $\chi=2$ at  lower  salt concentration of $n_0=20$~mM in Fig. \ref{f:neutral_pressure}d. Even though increasing $c_0$ up to $c_0=10$~mM enhances the attractive pressure as shown in the figure, the behavior of the effective interaction still remains in the weak-disorder regime as can be verified from the pressure components in Fig. \ref{f:kappa_c0}b (e.g., compare $P_{mon}$ and $P_c$ with those in Fig. \ref{f:neutral_pressure}e). This is clearly because of strong salt screening effects at large $n_0$, in which case the strong-disorder behavior can be achieved only by taking even larger values of $\chi$.

\section{Conclusion and Discussion}
\label{sec:conclusion}

In this work, we have studied the effective interactions between {\em net-neutral} dielectric slabs that carry {\em quenched} (fixed) random charge distributions on their apposing surfaces while they are immersed in an asymmetric aqueous electrolyte (Coulomb fluid) containing mobile monovalent and multivalent ions. The effective interaction between quenched, random charge distributions have been considered in a series of recent works \cite{Bing,klein2,kantor-disorder0,andelman-disorder,Miklavic,Ben-Yaakov-dis,Lukatsky1,Lukatsky2,Rabin,ali-rudi,rudiali,partial,disorder-PRL,jcp2010,pre2011,epje2012,jcp2012,book,jcp2014,sm2015}, and, while the role of Debye screening due to a weakly coupled (monovalent) salt solution on these interactions have been analyzed, the role of multivalent ions that couple strongly with surface charges has received much less attention \cite{ali-rudi,partial,jcp2014,sm2015}.

The interaction between (electro-) neutral dielectric slabs is traditionally described in terms of vdW or Casimir-type forces \cite{Parsegian2005,Ninham76} that are always attractive between identical bodies  in vacuum or in a polarizable medium such as water or an aqueous salt solution. In the DLVO context \cite{VO,Israelachvili,andelman-rev,Safranbook,holm}, the vdW attraction is counteracted by the mean-field electrostatic repulsion between like-charged surfaces, providing thus a mechanism for the stability of colloidal dispersions. In the case of strictly neutral surfaces, however, the electrostatic inter-surface repulsion is obviously absent but, in addition to the vdW interaction, there are image-induced, ion-depletion forces that are attractive as well and arise because the mobile solution ions are depleted from the vicinity of the  dielectric interfaces as a consequence of the repulsion from their image charges (these image charges are of the same sign when the surfaces are immersed in a medium of larger dielectric constant as compared with that of the slabs). This effect is intrinsically non-mean-field and occurs because of the discrete nature of ions, neglected in the collective mean-field description based on the standard Poisson-Boltzmann theory \cite{VO,Israelachvili,andelman-rev,Safranbook,holm}. Image-induced, ion-depletion effects for strictly neutral dielectric surfaces have been studied extensively in recent years  (see, e.g., Refs. \cite{Hansen,Dean04,Zemb,Kjellander87,Kjellander08,Netz1,Netz2,Netz3,Monica,Monica2,Monica3,Wang0,Wang1,Wang2,Levin1,Levin2,Levin2b,Levin3,Levin4,Lue0,Lue1,Lue1b,Lue2,Markovich1,Markovich2,Sahin0,Sahin5,Sahin2,Sahin3,Sahin4,SCdressed3} and references therein) and the role of multivalent ions, in particular, has been considered in Refs. \cite{SCdressed3,Kjellander08}.

In the present work, we add a new feature to the current understanding of the effective interaction between neutral surfaces and consider the situation, in which the surfaces are  neutral only {\em on the average} but otherwise carry a quenched random distribution of positive and negative charges. Indeed, heterogeneously or randomly charged surfaces are commonplace in soft matter and have attracted a lot of attention in recent years (see, e.g., Refs. \cite{kim2,kim3,barrett,science11,liu,Meyer,Meyer2,klein,klein1,klein2}). We have thus shown that the presence of {\em quenched surface charge disorder} in conjunction with {\em mobile multivalent ions} in the ionic solution leads to remarkable and significant qualitative changes in the standard picture commonly accepted for the interaction between neutral bodies.

The charge disorder has several manifestations in the present context: First, the self-interactions between the quenched random charges on the inner surfaces of the dielectric slabs and their image charges lead to a repulsive, short-ranged interaction pressure, $P_{dis}$, that tends to counteract the attractive vdW interaction pressure, $P_{vdW}$ as discussed thoroughly elsewhere \cite{rudiali,disorder-PRL,jcp2010,pre2011,epje2012,jcp2012,book}. 
The surface charge disorder, on the other hand, modifies the single-ion interaction potentials  in a way that creates a singular attractive potential acting between ions and the dielectric boundaries \cite{ali-rudi,partial,jcp2014,sm2015}. This effect works against the ion-image repulsions that tend to deplete ions from the slit region between the slabs. Thus, as a result of this attractive disorder-induced potential, more mono- and multivalent ions are accumulated in the slit when the confining dielectric boundaries are randomly charged. However, we find qualitative differences in the ways this excess attraction affects the osmotic pressure of ions. The osmotic pressure due to monovalent ions, $P_{mon}$, can become {\em repulsive} which, in fact, agrees with the standard mean-field picture since monovalent ions are only weakly coupled to the bounding surfaces and create a larger entropic pressure upon further accumulation in the slit. On the other hand, the osmotic pressure due to multivalent ions, $P_c$, can become even more {\em attractive}! This is in clear contrast with the image-induced, ion-depletion mechanism, in which a {\em decrease} in the number of mono- and multivalent ions in the slit gives rise to (stronger) attractive osmotic pressures on the slabs, similarly for both types of ions.

This rich behavior stems from combined effects of surface charge disorder and the presence of mobile multivalent ions, with the latter creating strong inter-surface attractions upon further accumulation in the slit. As a result, the effective total interaction pressure between randomly charged, net-neutral dielectric surfaces can differ qualitatively from what one expects based on the standard vdW and/or image-induced, ion-depletion interactions for strictly neutral surfaces, where they predict purely attractive interactions (unless at relatively large bulk salt concentrations, e.g., above 250mM, as noted in Ref.  \cite{Kjellander08}). In particular, the net effect from the competing electrostatic components due to disorder, mono- and multivalent ions results in a distinct, disorder-induced {\em non-monotonic} behavior. 

In the {\em weak disorder regime}, the net electrostatic pressure becomes gradually more repulsive (positive) at intermediate to large separations as the disorder coupling parameter (or equivalently the disorder variance, $g$) on the bounding surfaces of the slabs is increased from zero (strictly neutral or disorder-free system),  exhibiting first a shallow minimum and then a pronounced repulsive hump at intermediate separations. As the disorder coupling parameter is further increased into the {\em strong disorder regime} (typically beyond $\chi=2\pi q^2\lB^2g\simeq 1$), the behavior of the net electrostatic pressure is reversed; the repulsive hump is diminished and one finds a strongly attractive (negative) interaction pressure with a range and magnitude larger than that of the vdW or the image-induced, ion-depletion interaction pressures between strictly neutral surfaces. The crossover from weak to strong disorder regimes can be effected  also by decreasing the salt screening parameter (through decreasing the bulk concentration of monovalent ions) and/or increasing the bulk concentration of multivalent ions. The two different regimes of the disorder effects, the strong and the weak disorder regimes, bear also some resemblance to the strong-weak coupling dichotomy that exists for the net-charged surfaces, dependent on the electrostatic coupling parameter \cite{perspective}. It is no small feat to be able to partition the behavior of this complicated system by two dimensionless coupling parameters that effectively govern its salient characteristics.

Finally, we note the qualitative differences between the ion-mediated interactions we find between net-neutral surfaces in this work and those we reported between randomly charged surfaces carrying a finite mean surface charge density, $\sigma\neq 0$, in Refs. \cite{jcp2014,sm2015}. One of the key differences is that the repulsion between mean surface charges on the inner surfaces of the slabs, $P_{\sigma}$,  in the latter case contributes a dominant repulsive pressure to the net electrostatic pressure, which is stronger, both in magnitude and range, than $P_{dis}$ and $P_{vdW}$. On the other hand, when the surfaces carry a finite mean surface charge density, the osmotic pressure of monovalent ions, $P_{mon}$, turns out to be always repulsive and comparable in magnitude and range to $P_{\sigma}$; these repulsive pressure components completely mask the short-range interaction pressures, $P_{dis}$ and $P_{vdW}$, whose roles in competition with the attractive pressure due to multivalent ions, $P_c$, are brought up only in the case of net-neutral surfaces. Notably, we find that $P_{mon}$ in the case of surfaces carrying a finite mean surface charge density hardly responds to changes in the disorder strength, while, in the case of net-neutral surfaces, it can change from a purely attractive component (due to image-induced, ion-depletion effects) for disorder-free and weakly disordered surfaces to a repulsive one (due to disorder-driven ion accumulation in the slit) for strongly disordered surfaces.
Another difference between the two cases is that multivalent (counter-) ions in the case of surfaces carrying a finite mean surface charge density correlate strongly also with the mean charge densities of the confining boundaries of the slit, in line with the strong-coupling paradigm for electrostatics of charged surfaces known to generate strong like-charged surface attractions (see, e.g., Refs. \cite{Levin02,Shklovs02,hoda_review,Naji_PhysicaA,perspective,French-RMP}). Such strong-coupling effects are clearly absent in the present case with net-neutral surfaces.

In summary, our analysis provides new insight into the intricate role of surface charge disorder in the context of net-neutral surfaces and its fingerprints on the effective ion-mediated interactions between them. These interactions  are predicted to occur with a range and magnitude comparable to or, for strongly disordered systems, much larger than the vdW and/or image-induced, ion-depletion effects considered previously only in the case of strictly neutral surfaces. Our predictions will thus be amenable to numerical and/or experimental verification.

\section{Acknowledgments}

A.N. acknowledges partial support from the Royal Society, the Royal Academy of Engineering, and the British Academy (UK). H.K.M. acknowledges  support  from the School of Physics, Institute for Research in Fundamental Sciences (IPM), where she stayed as a short-term visiting researcher during the completion of this work.  R.P. acknowledges support from the Slovene Agency of Research and Development (ARRS) through Grant Nos. P1-0055 and N1―0019 as well as the travel grant from the Simons Foundation and the Aspen Center for Physics, supported by National Science Foundation grant PHY-1066293.


\begin{thebibliography}{99}
\bibitem{Note}
This constraint on multivalent counterions can be achieved by enclosing the outer regions `behind' the slabs in semi-permeable membranes, although, in the regime of parameters considered in this paper (e.g., with slab thicknesses much larger than the Debye screening length), this constraint has no sizable impact on our results.
\end{thebibliography}
\end{document}